\documentclass[final,5p,times,twocolumn,authoryear]{elsarticle}
\journal{Particuology}
\graphicspath{{./figures/}}

\usepackage[authoryear]{natbib} 
\usepackage[colorlinks]{hyperref}
\usepackage{doi}

\usepackage{graphicx,xcolor}
\usepackage{amssymb, amsmath,siunitx, bm}
\usepackage{nomencl}
\usepackage[capitalise]{cleveref}
\usepackage{booktabs}
\usepackage{nth}
\usepackage{enumitem}
\usepackage{float, subfigure}
\usepackage{diagbox}

\usepackage{lineno, todonotes, easyReview, csquotes}
\makenomenclature

\begin{document}
\begin{frontmatter}

	\title{Modelling of pressure drop in periodic square-bar packed beds} 
	\author[label1]{Hakan Demir\corref{cor1}}
	\ead{H.Demir@ruhr-uni-bochum.de}
	\cortext[cor1]{Corresponding author}
	\author[label1]{Wojciech Sadowski} 
	\author[label1]{Francesca di Mare} 
	\affiliation[label1]{organization={Ruhr University Bochum},
		addressline={Department of Mechanical Engineering},
		city={Bochum},
		postcode={44801},
		state={North Rhine-Westphalia},
		country={Germany}}

	\begin{abstract}
    Understanding fluid flow through porous media with complex geometries is essential
    for improving the design and operation of packed-bed reactors. Most existing 
    studies focus on spherical packings,
    having as a consequence that accurate models for irregular interstitial geometries are scarce.
    In this study, we numerically investigated the flow through a set of packed-bed geometries 
    consisting of square bars stacked on top of each other and arranged in disk-shaped modules.
    Rotation of each module allows the generation of a variety of geometrical configurations 
    at Reynolds numbers of up to 200 (based on the bar size). 
    Simulations were carried out using the open-source solver OpenFOAM.
    Selected cases (e.g., $\alpha = \SI{30}{\degree}$, $\mathrm{Re}_\mathrm{p} = 100, 200$)
    were compared against Particle Image Velocimetry measurements. Results reveal that,
    based on the relative rotation angle, the realized geometries can be classified as
    channel-like ($\alpha \leq \SI{10}{\degree}$)
    and lattice-like ($\alpha \geq \SI{15}{\degree}$),
    fundamentally altering the friction factor. Furthermore,
    the maximum friction factor obtained in the creeping regime occurred at
    $\alpha = \SI{25}{\degree}$, whereas in the inertial regime, this
    occurred at $\alpha = \SI{60}{\degree}$. The module-equivalent diameter,
    based on the angle-dependent wetted surface area, collapses the friction factor
    onto the Ergun correlation and yields good permeability predictions for the
    lattice-like geometries.
	\end{abstract}


\end{frontmatter}

\section{Introduction}

A packed bed consists of a solid packing, typically a particle assembly
confined within an enclosure, and passed by a fluid.  Packed beds are widely
used in petroleum, process, and  chemical industry \citep{seckendorff21}.
Typical examples are  catalyst reactors \citep{MiguelGarcia20}, where residual
methane undergoes total oxidation to produce carbon dioxide and water;
catalytic converters for emission control, e.g. \citep{Schnitzlein87}, and
thermal energy storage systems \citep{RyanAnderson15}, or as filters in
environmental engineering \citep{ChenWang19}. The solid  particles are often
modelled as monodisperse spheres; in reality, however, they can span wide
ranges of sizes and shapes, e.g., cylinders, polyhedra, or arbitrary
non-polyhedral forms such as in \citep{Afandizadeh2001, Schlipf2015}.

The presence of the solid phase forces the fluid, in general, to adopt complex
meandering paths as it permeates the packing, enhancing the heat and mass
transfer and facilitating reactions \citep{Dixon20}. The gas flow in the
interstitial space, which is the focus of this study, is influenced by both
micro- and macroscopic parameters of the packing, e.g., particle size and shape
(or pore size), bed porosity and permeability, and the operating conditions of
the device characterized for example, by the Reynolds number \citep{Kumar23,
	Abdulmohsin17}. Therefore, a clear understanding of the impact of these factors
on the flow features is crucial for designing efficient reactors
\citep{Elzubeir24}.

Due to the complexity of these systems, detailed investigation,  both numerical
and experimental, covering all relevant scales is not viable. Therefore,
numerical models of varying fidelity are generally used in the design process.
For example, the whole device can be represented with a fast, one-dimensional
model  neglecting the geometrical complexity \citep[][e.g.,]{niedermeier2018,
	Nash2017}. When the problem has to be modelled in two or three dimensions, the
flow field can be described at a macroscale level (i.e., using averaged  flow
properties) using porous medium models \citep{collazo2012, battiato2019,
	Sadowski2023}. This typically  involves modelling the drag induced by the porous
medium using  closures \citep{Whitaker1986,whitaker1996} such as
Darcy–Forchheimer (DF). Complementary to these  continuum approaches, particle
movement and macroscopic  parameters of the particle assembly can be tracked by
the Discrete Element Method \citep{Cundall1979,illanamahiques2023,Ma2022}.

Higher accuracy of representation of the flow field is possible resolving the
geometry of the packing, and ensuring a sufficiently accurate spatio-temporal
discretisation to capture all relevant scales \citep{Dixon20}. Such
approach---called Particle-Resolved Direct Numerical Simulation (PRDNS)---offers
the most precise representation of both micro- and macroscale flow properties
within packed beds.  PRDNS can provide detailed data on interphase momentum,
energy, and mass transfer allowing to develop, calibrate, and validate closure
models for unresolved simulations
\citep{Tenneti2014,Sadowski2023b,Sadowski2023}. However, such simulation
requires high computational effort and ,hence, is typically limited to simple
and laboratory-scale configurations \citep{neeraj2023,sadowski24,sadowski2025}.

A further current limitation of investigations of particle assemblies and packed
beds is the simplification of the particle geometry, whereby particles are
generally considered as spherical \citep[e.g.,][]{zhu16, dentz22, storm24}, even
in approximations of industrial devices \citep{neeraj2023, sadowski24}, for
example in a body-centered  cubic (BCC) arrangement. Spherical particles offer
practical advantages  as the calculation of wall-distances can be enormously
simplified and  the definition of particle contact regions is relatively
straightforward.  This is more challenging for irregular shapes
\citep{Dixon20,jurtz2019}  so that spherical particles allow for efficient
contact  detection and simulation of full-scale systems
\citep{illanamahiques2023a}.

However, the higher complexity of the flow field patterns induced by
heterogeneous, non-spherical packings can result in large errors in velocity
predictions by models describing the flows in terms of averaged velocity
\citep{Moghaddam19}. Moreover, \citet{roding2017} observed that the effective
diffusivity in random packings of cuboids, spheres and ellipsoids is
significantly influenced by the particle shape at identical porosity. This is
troubling from the perspective of unresolved flow modelling, as most models
express the drag or diffusivity in terms of macroscopic parameters and would
be unable to capture such differences. Standard correlations for pressure drop,
such as the Ergun equation \citep{ergun1951} or Kozeny--Carman model
\citep{carman1937}, were originally developed only for spherical particles.
These and similar drag laws have, to some extent, been successfully generalized
to other geometries \citep[e.g.,][]{liu94,duplessis1988,duplessis2008}
and adjusted to take into account the presence of other particle shapes,
	either by including the sphericity of the particles as a parameter of the
	model \citep{woudberg2020,Li2011EffectiveDiameter}, or by constructing a multiscale model
	accounting for average anisotropy of the packing based on microscale
	simulations \citep{tobis2008,tobis2000}. That said, to the best of our
	knowledge, the influence on the induced drag by the variation in packing
	geometry at a constant macroscopic geometrical properties (e.g, porosity) remains largely unquantified.

The geometry investigated in this study is based on the modular  packed-bed
reactor designed by \citet{christin2025}, which consists of disk-shaped modules
containing square bars that can be rotated relative to one another. This allows
for the study of a wide range of interstitial configurations can be studied,
while  keeping the porosity, bed cross-section and the shape of modules strictly
constant. Although this geometry does not represent a specific industrial
		packing, it provides  a controlled framework  for studying flow through complex
		interstitial geometries. Crucially, the availability of detailed Particle Image
		Velocimetry (PIV) measurements \citep{christin2025} enables direct validation of
		the numerical results, which is rarely possible for non-spherical packed beds.

Therefore, this study explores how the variation of the interstitial geometry
affects the induced drag and the permeability in a dense packed bed consisting
of square bars. To quantify these effects, we carried out numerical simulations
in a set of different geometries with the same porosity ($\phi=0.322$) at
Reynolds numbers (based on bulk velocity and bar size) spanning the range
between $0.1$ and $200$. Special attention was given to the influence of the
geometry on the transition  from viscosity- to inertia-dominated flow regimes.
Moreover, to ensure the quality of the comprehensive database resulting from
the 266 simulations, we carried out a detailed validation of the numerical
setup against the PIV results of \citet{christin2025}. Furthermore, a thorough
mesh sensitivity analysis was performed to assess the viability of the
numerical approach.

The remaining part of this work is organized as follows. In \Cref{sec:geometry},
the geometry of the packed bed and the associated flow conditions are described
in detail.  In \Cref{sec:math,sec:num} the mathematical and numerical models
used in the study, respectively, are described. \Cref{sec:results} presents the
results with particular emphasis on the non-dimensional parameters in porous
media.  Finally, we summarize our conclusion in \Cref{sec:Conclusion}.

\begin{figure*}[tb]
	\def\FH{4.0cm}
	\centering
	\subfigure[]{
		\includegraphics[height=\FH]{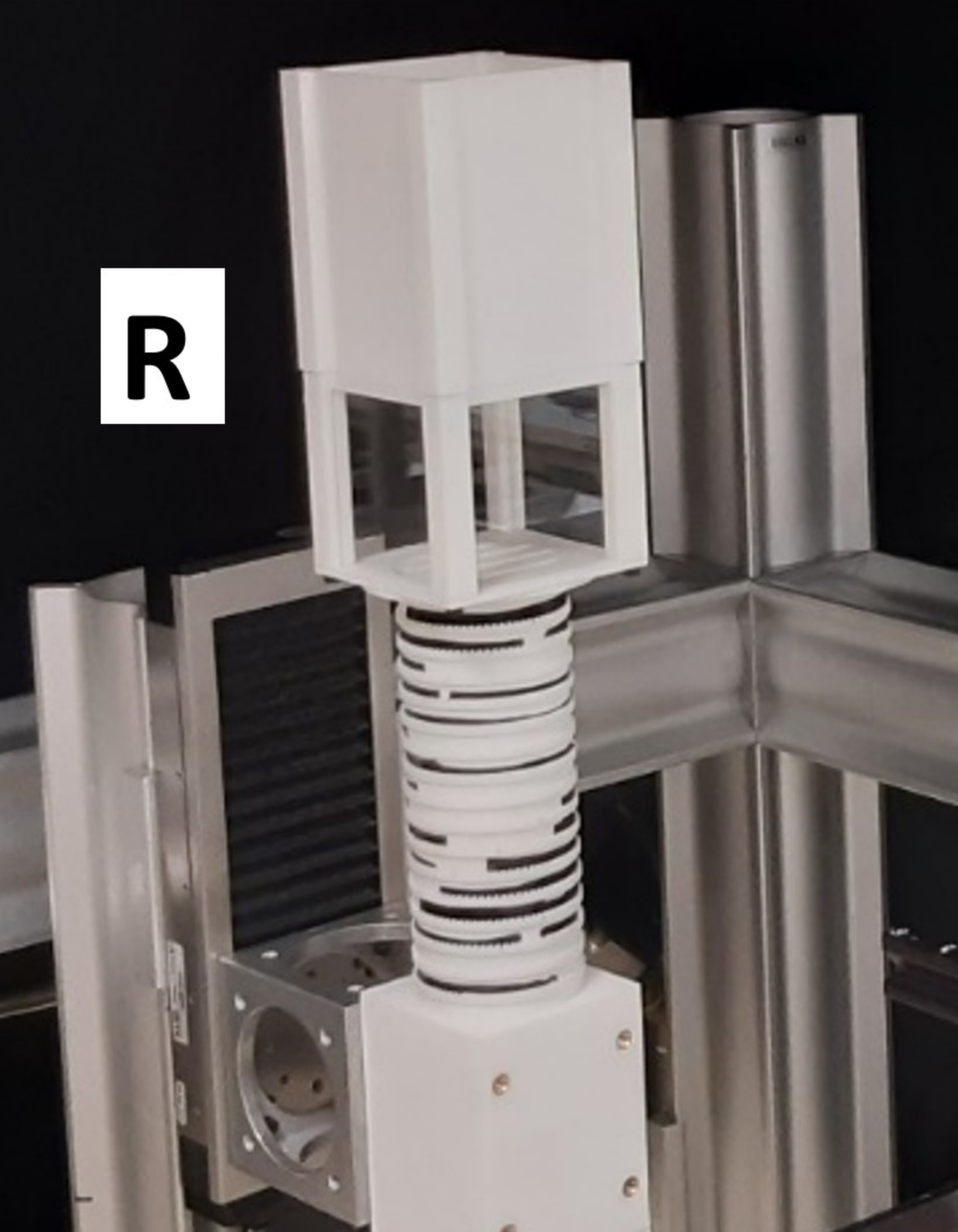}
		\label{fig:reactor}
	}
	\hspace{0.02\textwidth}
	\subfigure[]{
		\includegraphics[height=\FH]{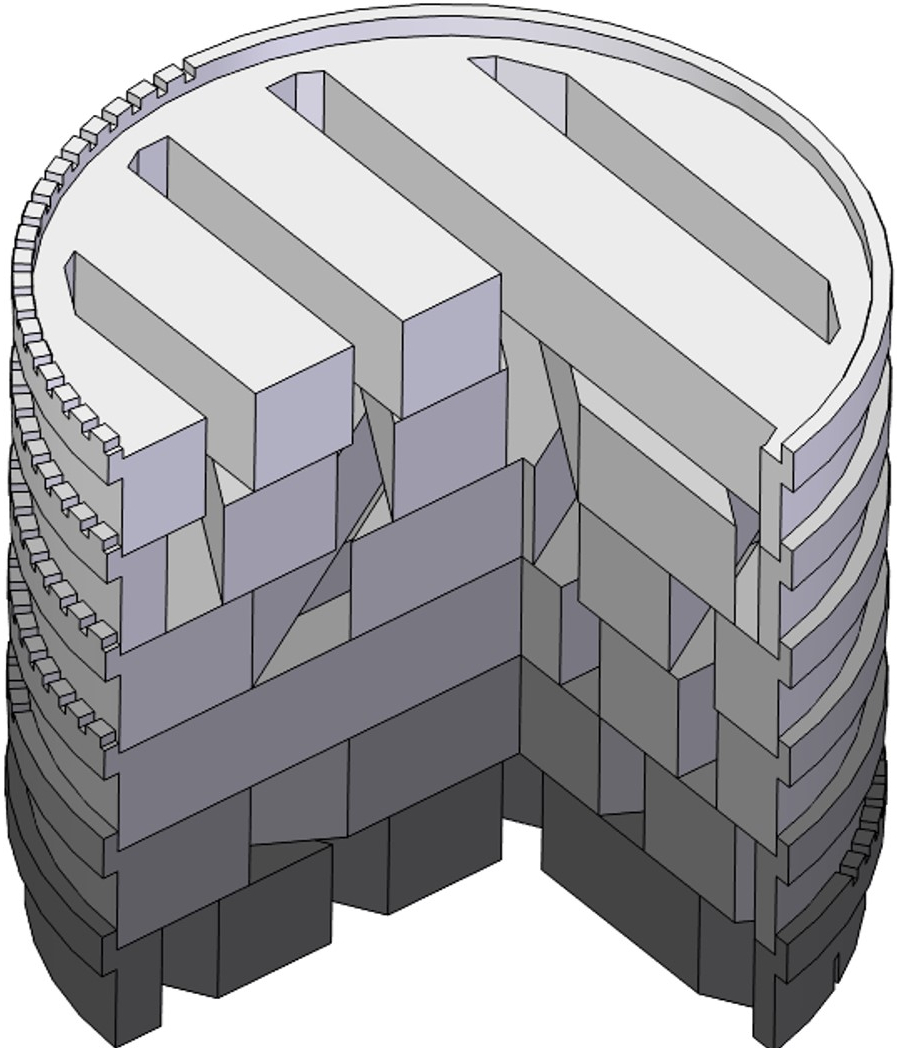}
		\label{fig:crossVoidSpace}
	}
	\hspace{0.02\textwidth}
	\subfigure[]{
		\includegraphics[height=\FH]{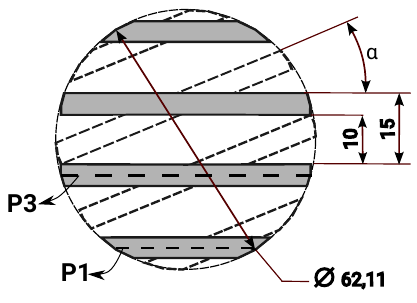}
		\label{fig:ModuleGeometry}
	}

	\caption{(a) The laboratory-scale packed bed of \citet{christin2025}. (b) The
		cross-section of the packed-bed geometry \citep{christin2025} formed from six modules, each
		rotated by \SI{30}{\degree} relative to the preceding one. (c) Schematic of
		the geometry of each module showing the circle enclosing the dodecagon
		defining the outer geometry of the slits. The streamwise direction is
		oriented perpendicular to the surface of the paper.}
	\label{fig:geometryOverview}
	\undef\FH
\end{figure*}

\begin{figure}[tb] \centering
	\includegraphics[width=0.65\linewidth]{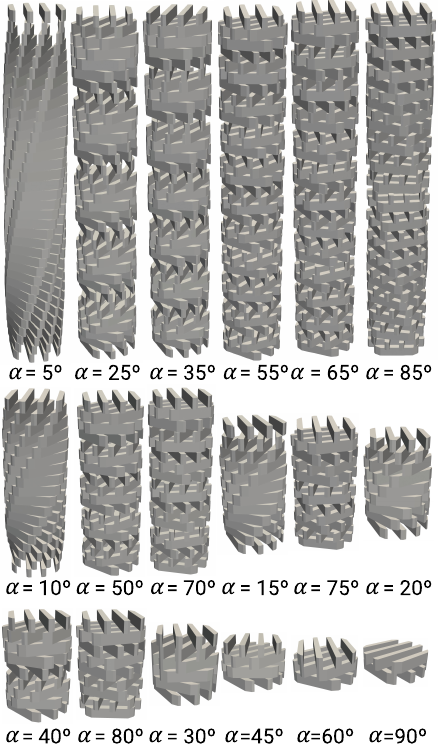} 
	\caption{Visualizations of the simulation domains (i.e.,
		volumes occupied by the fluid) for each of the studied rotation angles,
		ranging from $5^\circ$ to $90^\circ$. The streamwise direction is oriented
		vertically, with the flow moving from bottom to top.}
	\label{fig:base_geometry}
\end{figure}

\section{Geometry and flow conditions}\label{sec:geometry}
The geometry investigated in this work is based on a laboratory-scale modular
packed-bed reactor designed by \citet{christin2025}. The reactor, shown in
\cref{fig:reactor}, is a column consisting of identical modules and an outlet
section which is not considered in this study. Each module is a disk with a
thickness of $B = \SI{10}{mm}$, with \SI{5}{mm} wide slits cut through, forming
prismatic bars with a square cross-section ($B \times B$) within each module.
The corresponding dimensions are shown in \cref{fig:ModuleGeometry}. The
geometry of the void space in each module is defined so that the side walls of
the slits are lying on a regular dodecagon inscribed in a circle with a
diameter $D = \SI{62.12}{mm}$, which is concentric with the modules outer
geometry. The area of this circle defines the bed’s cross-sectional area,
leading to a theoretical porosity of $\phi = 0.322$.

The disks are stacked vertically and rotated relative to one another, creating a
complex void-space geometry between adjacent layers. An example of the resulting
geometry, with a rotation angle of $\SI{30}{\degree}$ between layers, is shown
in \cref{fig:crossVoidSpace}.

This geometry was designed to allow for various geometries of interstitial
space: the rotation angle, height of the modules or the size of the bars can be
adjusted independently. While not directly tied to a specific industrial
application, the use of bars in each module enables spatio-temporally resolved
optical measurements in each configuration \citep{christin2025}, unlike
typically encountered, random non-spherical packing. This last property makes
it valuable for studying flow characteristics \textit{via} simulation
approaches, as the available measurement data allows for the detailed
validation at the pore scale.

By rotating each module by a fixed angle, different interstitial configurations
and flow characteristics in the bed can be investigated. In the present study,
\num{19} rotation angles $\alpha = \SI{0}{\degree}, \SI{5}{\degree}, \dotsc,
	\SI{90}{\degree}$ and, thereby, 19 geometries, are considered. In order to
accurately represent the macroscopic flow properties of each of packing, it is
vital that a periodic geometry is constructed \citep{guibert2016,scandelli2022},
otherwise a blockage effects would influence the computed values of pressure
drop. For this reason, for each rotation angle, such a number of layers is
chosen, which results in a periodically repeating geometry. The number of
layers corresponding to each rotation angle is provided in
\cref{tab:NumberOfCellsAndLayers}.

\begin{figure}[tb]
	\centering
	\includegraphics[width=0.45\textwidth]{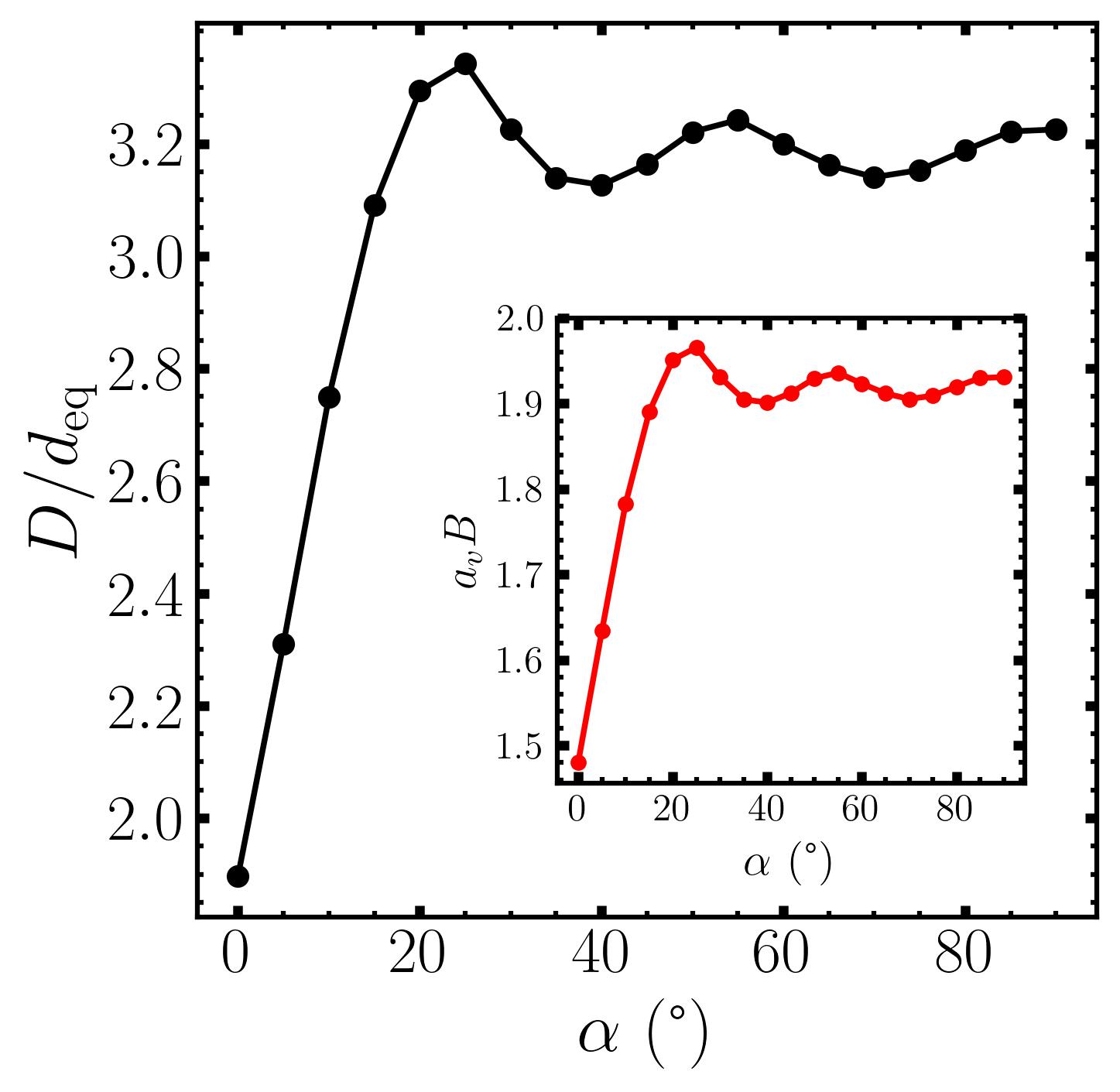}
	\caption{Non-dimensional specific surface area~$a_v B$ and bed-to-particle diameter ratio~$D/d_{\mathrm{eq}}$
	as a function of the rotation angle~$\alpha$}
	\label{fig:D_dptubetobedratio}
\end{figure}

\Cref{fig:base_geometry} shows the simulation domains corresponding to each
angle.
The studied geometries can be classified by specific surface area $a_v$ defined
as the ratio of the area of the wetted surface to the total volume of the
packed bed, and the ratio of the
diameter of the bed to the equivalent particle diameter $D/d_\mathrm{eq}$, both plotted in \cref{fig:D_dptubetobedratio} (see
\cref{sec:math} for detailed discussion of evaluating both parameters for a given geometry).

At small angles the bars in adjacent modules are nearly aligned, creating
large areas of contact between the bars and yield the
lowest values of~$a_v$. As~$\alpha$
increases, more surface is exposed, so
that~$a_v B$ rises to about $2.0$ near
$\alpha = 25^{\circ}$. For $\alpha \geq 15^{\circ}$,~$a_v B$
oscillates with~$\alpha$, reflecting the
periodically changing contact pattern between successive layers.
Since
$d_{\mathrm{eq}} \propto 1/a_v$
(\cref{eq:d_eq,eq:sauter-mean-diam}), $D/d_{\mathrm{eq}}$ follows the same
trend, ranging from approximately $1.9$ at $\alpha = 0^{\circ}$ to
$3.1$--$3.3$ for $\alpha \geq 15^{\circ}$.

For the current geometry, we picked $D/d_{\mathrm{eq}} \approx 3$ to marks the boundary
between the channel-like ($D/d_{\mathrm{eq}} < 3$) and lattice-like
($D/d_{\mathrm{eq}} > 3$) geometries, with the transition occurring between
$\alpha = 10^{\circ}$ and $15^{\circ}$.
Therefore, for small rotation angles ($\alpha \leq \SI{10}{\degree}$), the
arrangement resembles a \emph{channel-like} structure, where aligned void spaces
create curved flow passages with minimal obstruction. As $\alpha$ increases
beyond \SI{10}{\degree}, the void spaces become increasingly interconnected and
the geometry transitions to a \emph{lattice-like} structure with multiply
connected flow paths.
The \SI{90}{\degree} configuration, with
$D/d_{\mathrm{eq}} = 3.23$ and $a_v = 0.193~\mathrm{mm^{-1}}$, is
classified within the lattice-like group based on these geometric
parameters (see \cref{sec:results}).

These values of $D/d_\mathrm{eq}$ place the RC2 geometry in the low bed-to-particle diameter ratio regime
($D/d < 5$) \citep{yang2016}, comparable to the random packings of spheres
and cylinders studied by \citet{Moghaddam19} ($N = 2.29$--$6.1$) and
\citet{papkov2024} ($D/d = 2.9$ and $4.8$).

The fluid moves through the column in the axial direction, with the flow
conditions characterized by the pore Reynolds number
The pore Reynolds number is defined as
\begin{equation}
	\mathrm{Re}_\mathrm{p} = \frac{\langle w \rangle  B}{\nu},
	\label{eq:reynoldsnumber}
\end{equation}
where $B$ corresponds to the bar size,
$\langle w \rangle$ is the intrinsic average of the streamwise
velocity,
and $\nu$ is the kinematic viscosity.
(In the experimental setup of \citep{christin2025}, the intrinsic
average of velocity is computed directly from the volumetric flow rate forced
through the packed bed.)

		Depending on the application, the flows in packed beds and porous materials
		can be characterized by a wide range of Reynolds numbers
		\citep{wood2020,Dullien1975}, requiring the accurate description of both
		the viscous and inertial effects, to properly model the
		flow behaviour. Therefore, in the present paper we focus on a Range of
		Reynolds numbers spanning $0.1$ to $200$.
	
		According to the flow regime
		classification of \citep{Dybbs1984ANL}, the transition to turbulence is
		typically reported at the pore-size Reynolds numbers in the range
		$\mathrm{Re}_p \approx 300$--$500$ for packed beds of spheres.
		\citep{wood2020}, in their review of turbulent flows in porous media,
		confirmed that these thresholds are broadly consistent across different
		packing configurations, although the exact transition depends on the
		specific geometry, and well-defined turbulence can be established above
		$\mathrm{Re}_p\sim750$. Since the present study is limited to $Re_p \leq
			200$, all considered flow conditions can be expected to be laminar, with,
		for higher Reynolds numbers, strong inertial effects and unsteady behaviour
		which can exhibit intermittency.

\section{Mathematical  model}\label{sec:math}

An incompressible flow of a Newtonian fluid with constant and homogeneous
density (i.e., neglecting variations due to mixing, reaction, or thermal
effects) inside the considered packed bed can be described at the pore
level by the Navier–Stokes and continuity equations:

\begin{subequations}
	\begin{equation}
		\frac{\partial u_i}{\partial x_i} = 0,
		\label{eq:MassConservation}
	\end{equation}
	\begin{equation}
		\frac{\partial u_i}{\partial t}
		+ u_j \frac{\partial u_i}{\partial x_j} = -\frac{\partial p}{\partial x_i}
		+ \nu \frac{\partial^2 u_i}{\partial x_j \partial x_j},
		\label{eq:momentumEquation}
	\end{equation}
\end{subequations}
where $p$ represents the kinematic pressure (i.e., static pressure divided by the
fluid density) and $\bm{u} = [u, v, w]^T$ is the velocity.

To describe the flow at the macroscale, governing equations formulated in terms
of averaged flow properties are necessary. The initial work in this field was
undertaken by \citet{darcy1856}, who established a linear relationship between
the pressure drop and the velocity through a porous bed, under the
creeping flow regime. When inertial effects become significant, the
relationship between pressure gradient and velocity is no longer linear. A
corrected equation accounting for fluid inertia by introducing a term
proportional to the square of the velocity was presented by \citet{Forchheimer1901}
and is known as the Darcy–Forchheimer equation for the pressure drop.

Although the DF model was originally postulated as empirical correlations,
it has been successfully rederived by upscaling the equations valid at the
pore scale \citep{Whitaker1986,whitaker1996}. Most of these derivations fundamentally rely on some form of
averaging technique, such as the Volume Averaging Method (VAM) \citep{battiato2019}.
In the framework of VAM, an intrinsic average of a flow variable inside a
porous medium is defined as
\begin{equation}
	\langle \psi \rangle = \frac{1}{V_f}\int\limits_{V_f} \psi \text{d}\,V,
	\label{eq:intrinsic-avg}
\end{equation}
and is connected to a superficial average ${\langle \psi \rangle}_s$ by means of the
Dupuit's relation ${\langle \psi \rangle}_s = \phi{\langle \psi \rangle}$.
A uniform flow in the vertical direction (along the $z$ coordinate axis) in the
packed bed can be described as \citep{woudberg2020}:
\begin{equation}
	\frac{\partial \langle p \rangle}{\partial z} =
	-\frac{\nu}{K_\mathrm{eff}}\langle w \rangle_s
	\approx
	-\frac{\nu}{K}\left( 1 + C_\mathrm{F} \frac{\phi\sqrt{K}}{B}\mathrm{Re}_\mathrm{p} \right) \langle w \rangle_s,
	\label{eq:DarcyForchheimerEquation}
\end{equation}
where $K_\mathrm{eff} = K_\mathrm{eff}(\mathrm{Re}_\mathrm{p})$ is the
effective permeability of a porous material which accounts for both viscous and
inertial effects, and is, therefore, a function of the Reynolds number. It can
be approximated with the DF drag model formulated in terms of the isotropic
permeability $K$, which is a limit of $K_\mathrm{eff}$ at zero
		Reynolds number, and the Forchheimer (or inertial) coefficient $C_\mathrm{F}$.
The values of $K$ and $C_\mathrm{F}$ can be determined experimentally, computed
from simulations, or estimated using semi-empirical correlations
\citep[e.g.,][]{Lenci2022,woudberg2020, liu2024a}. The effective permeability
can be non-dimensionalised by selecting a characteristic length scale of the
studied problem, here $B$, and forming the Darcy number
$\mathrm{Da}_{\mathrm{(eff)}} = K_{\mathrm{(eff)}}/B^2$:
\begin{equation}
	\frac{K_\mathrm{eff}}{B^2} \approx \frac{\mathrm{Da}}{1 + C_\mathrm{F} \phi \sqrt{\mathrm{Da}} \mathrm{Re}_\mathrm{p}}.
	\label{eq:DarcyForchheimarpermeability}
\end{equation}
It is also useful to express the pressure drop in terms of a friction factor, as commonly done in the packed-bed literature \citep{Dullien1975}.
The
pressure gradient can be normalized by the characteristic dimension of the
studied geometry, taken here again as $B$, and the square of the superficial
velocity:
\begin{equation}
	f_\mathrm{p} = -\frac{B}{\langle w \rangle_s^2} \frac{\partial \langle p \rangle}{\partial z},
	\label{eq:friction_factor}
\end{equation}
which combined with \cref{eq:DarcyForchheimarpermeability} gives:
\begin{equation}
	f_\mathrm{p} \approx \frac{1 + C_\mathrm{F} \phi \sqrt{\mathrm{Da}}\mathrm{Re}_\mathrm{p}}{\phi \mathrm{Re}_\mathrm{p} \mathrm{Da}}.
	\label{eq:friction_factor_nondimensional}
\end{equation}
Importantly, the choice of characteristic dimension is not unique and other
definitions of the friction factor have been proposed. A common approach uses
the specific surface area $a_v=A_p/V_p$ of a particle to define the Sauter-mean
diameter
\begin{equation}
	d_\mathrm{sd} = \frac{6}{a_v} = \frac{6\, V_p}{A_p}\,.
	\label{eq:sauter-mean-diam}
\end{equation}
where $V_p$ and $A_p$ are the
volume and surface area of the particle, respectively.
The Sauter-mean diameter does not account for particle non-sphericity.
To incorporate shape effects, the sphericity $\psi$ \citep{Wadell1935}
is introduced as the ratio of the surface area of a volume-equivalent sphere
to that of the actual particle:
\begin{equation}
	\psi =
	\frac{\pi^{1/3} \left( 6\, V_{\mathrm{p}} \right)^{2/3}}
	{A_{\mathrm{p}}}\,.
	\label{eq:sphericity}
\end{equation}
\citet{Li2011EffectiveDiameter} proposed the equivalent particle
diameter, defined as
\begin{equation}
	d_\mathrm{eq} = \psi\, d_\mathrm{sd}\,,
	\label{eq:equiv-diam}
\end{equation}
which reduces to $d_\mathrm{sd}$ for spherical particles where
$\psi = 1$.
Using such length scale, the \citet{ergun1951} equation can be used to model the
drag in packed beds of non-spherical particles \citep{li2011}, which
is often expressed in the form of the following friction factor:
\begin{equation}
	f_\mathrm{Erg} = -\frac{\partial \langle p \rangle}{\partial z}
	\frac{d_\mathrm{eq}}{\langle w \rangle_s^2}
	= \frac{1-\phi}{\phi^3}\left(\frac{150}{\mathrm{Re}_\mathrm{Erg}} + 1.75\right),
	\label{eq:friction_factor_Erg}
\end{equation}
with the Reynolds number defined as
\begin{equation}
	\mathrm{Re}_\mathrm{Erg} = \frac{\langle w \rangle d_\mathrm{eq}}{\nu} \frac{\phi}{1-\phi}.
	\label{eq:reynoldsnumberErg}
\end{equation}
\Cref{eq:friction_factor_Erg} can be recast in the form equivalent to
\cref{eq:DarcyForchheimerEquation} with
the Blake--Kozeny model for permeability
\begin{equation}
	K^\mathrm{Erg} = \frac{d_\mathrm{eq}^2}{150}\frac{\phi^3}{(1-\phi)^2}
	\label{eq:K-ergun}
\end{equation}
and the Forchheimer coefficient expressed as:
\begin{equation}
	C^\mathrm{Erg}_\mathrm{F}
	= \frac{1.75\,(1-\phi)\,\sqrt{K/B^2}}{\phi^{3}}.
	\label{eq:CF-ergun}
\end{equation}

The definitions of the friction factor fundamentally depend on how the
characteristic dimension is computed, as it influences both the definition of
the Reynolds number and the coefficients of drag. Because the RC2 module
contains interconnected bars rather than loose particles, two definitions of
characteristic diameters can be considered: one based on the module geometry as
a whole, and one based on a single representative bar. These are evaluated in the
following section.

\subsection{Geometric parameters for the RC2 configuration}

In the present work, the specific surface area is primarily defined as
\begin{equation}
	a_v = \frac{A_{\mathrm{wetted}}}{V_{\mathrm{total}}}\,,
	\label{eq:av-def}
\end{equation}
where $A_{\mathrm{wetted}}$ is the total wetted surface area of the module
and $V_{\mathrm{total}}$ is the total module volume (comprising both fluid
and solid regions). Because the bar-to-bar contact regions change with the
rotation angle~$\alpha$, so does the wetted surface exposed to the fluid,
and consequently both~$a_v$ and the bed-to-particle diameter ratio
$D/d_{\mathrm{eq}}$ vary with~$\alpha$
(\cref{fig:D_dptubetobedratio}).
The equivalent diameter $d_{\mathrm{eq}}$ is derived from the module as a whole.
The module-based sphericity is
\begin{equation}
	\psi_{\mathrm{eq}}
	= \frac{\pi^{1/3}\,\bigl(6\,V_{\mathrm{p}}\bigr)^{2/3}}
	{A_{\mathrm{wetted}}}\,,
	\label{eq:psieq}
\end{equation}
The Sauter-mean diameter follows directly from \cref{eq:sauter-mean-diam}
and likewise depends on~$\alpha$.
The module-based equivalent diameter then follows from
\cref{eq:equiv-diam,eq:sauter-mean-diam} as
\begin{equation}
	d_{\mathrm{eq}} = \psi_{\mathrm{eq}}\,d_{\mathrm{sd}}^{\mathrm{eq}}\,.
	\label{eq:d_eq}
\end{equation}
%
One can also propose a characteristic length scale by establishing a
representative particle and use it to compute geometric properties: based
on a single representative square bar, we define the single-bar diameter
$d_{\mathrm{sb}}$. The simulation domain per module contains four square bars
arranged in two symmetrically positioned pairs that share the same
cross-section ($10\times10\;\mathrm{mm^{2}}$) but differ in length
($L_{1}=62\;\mathrm{mm}$ and $L_{2}=46\;\mathrm{mm}$).  A representative bar is
therefore defined with the averaged length $L_{\mathrm{avg}} = (L_{1}+L_{2})/2
	= 54\;\mathrm{mm}$, giving $V_{\mathrm{sb}} = 10\times10\times54 =
	5400\;\mathrm{mm^{3}}$ and $A_{\mathrm{sb}} = 2\,(10\times10) + 4\,(10\times54)
	= 2360\;\mathrm{mm^{2}}$.
The single-bar sphericity is
\begin{equation}
	\psi_{\mathrm{sb}}
	= \frac{\pi^{1/3}\,\bigl(6\,V_{\mathrm{sb}}\bigr)^{2/3}}
	{A_{\mathrm{sb}}}\,,
	\label{eq:psisb}
\end{equation}
where $V_\mathrm{sb}$ and $A_\mathrm{sb}$ are the volume and surface area of the representative square bar, respectively. For the representative bar described above, $\psi_{\mathrm{sb}} = 0.63$.
The Sauter-mean diameter for a single bar is
\begin{equation}
	d_{\mathrm{sd}}^{\mathrm{sb}}  = \frac{6\, V_\mathrm{sb}}{A_\mathrm{sb}}\,.
	\label{eq:sauter-mean-diamsb}
\end{equation}
The corresponding single-bar equivalent diameter reads
\begin{equation}
	d_{\mathrm{sb}} = \psi_{\mathrm{sb}}\,d_{\mathrm{sd}}^{\mathrm{sb}}\,.
	\label{eq:d_sb}
\end{equation}
Note that $d_{\mathrm{sb}}$ is constant for all~$\alpha$, since both
$\psi_{\mathrm{sb}}$ and $d_{\mathrm{sd}}^{\mathrm{sb}}$ are determined by the
fixed bar geometry, yielding $d_{\mathrm{sb}} = 8.65\;\mathrm{mm}$.
%
%
The Blake--Kozeny permeability estimate (\cref{eq:K-ergun}) can be evaluated
with each characteristic diameter:
\begin{equation}
	K_{\mathrm{eq}}
	= \frac{d_{\mathrm{eq}}^{2}}{150}\,
	\frac{\phi^{3}}{(1-\phi)^{2}}\,,
	\qquad
	K_{\mathrm{sb}}
	= \frac{d_{\mathrm{sb}}^{2}}{150}\,
	\frac{\phi^{3}}{(1-\phi)^{2}}\,.
	\label{eq:K-both}
\end{equation}
For both diameter definitions, the Ergun-type Reynolds number
(\cref{eq:reynoldsnumberErg}) is formed with the respective equivalent
diameter:
\begin{equation}
	\mathrm{Re}_{\mathrm{Erg}}^{d_{\mathrm{eq}}}
	= \frac{\langle w \rangle\, d_{\mathrm{eq}}}{\nu}\,
	\frac{\phi}{1-\phi}\,,
	\qquad
	\mathrm{Re}_{\mathrm{Erg}}^{d_{\mathrm{sb}}}
	= \frac{\langle w \rangle\, d_{\mathrm{sb}}}{\nu}\,
	\frac{\phi}{1-\phi}\,.
	\label{eq:Re-erg-both}
\end{equation}

\subsection{Tortuosity-based permeability model}

	The Blake--Kozeny and Kozeny--Carman models can be both expressed in a
	general sense as
	\begin{equation}
		K = \frac{\phi^{3}}{c_{\mathrm{KC}}\,\tau^{2}\,a_v^{2}}\,,
		\label{eq:K-tau}
	\end{equation}
	where $c_{\mathrm{KC}}$ is the Kozeny constant (for a cylindrical capillary
	$c_{\mathrm{KC}} = 2$) and $\tau$ is the Hydraulic tortuosity which quantifies
	the elongation of fluid paths relative to the straight-line distance through
	the medium. For spheres, $\tau$ is typically fixed at $\tau = \sqrt{2}$
	\citep{clennell1997,carman1937} leading to the Blake--Kozeny model.

	Naturally, there is no certainty that a model formulated with a constant
	value of $\tau$ will be valid across many geometries and estimating permeability
	by the use of \cref{eq:K-tau} may yield improved results.
	Different definitions of
	tortuosity have been proposed in the literature
	\citep{bear1972,clennell1997,matyka2008,duda2011}. Following
	\citet{duda2011}, the upper bound for the hydraulic tortuosity can be computed directly from
	the velocity field as the ratio of the volume-averaged velocity magnitude
	to the volume-averaged axial velocity component in the void space:
	\begin{equation}
		\tau_{\mathrm{eff}} = \frac{\langle |\bm{u}| \rangle}{\langle w \rangle}.
		\label{eq:tortuosity-eff}
	\end{equation}
	At the larger Reynolds numbers, the upper bound for the hydraulic tortuosity
	depends on $\mathrm{Re}_\mathrm{p}$ through the rearrangement of the flow
	field by inertial effects and is denoted~$\tau_{\mathrm{eff}}$. In the
	Darcy limit, it reduces to
	\begin{equation}
		\tau = \lim_{\mathrm{Re}_\mathrm{p} \to 0} \tau_{\mathrm{eff}}\,,
		\label{eq:tortuosity}
	\end{equation}
	which is a purely geometric property of the void space.

\section{Numerical  model}\label{sec:num}

All simulations were performed using OpenFOAM-12 \citep{greenshields2024}, an open-source finite
volume-based Computational Fluid Dynamics (CFD) software. Both steady-state and transient solvers were used,
employing the SIMPLE and PISO pressure-velocity coupling methods, respectively
\citep{moukalled2016}. To determine the appropriate solution method for each
geometry and flow regime, both algorithms were tested to compute the flow at $\mathrm{Re}_\mathrm{p} = 50$
in each geometry. The results showed no significant differences in the microscopic and
macroscopic flow features, such as the pressure gradient and the
intrinsic average of velocity
indicating that the steady-state solver is an appropriate choice for this
Reynolds number. Expecting the onset of unsteady flow for higher Reynolds
numbers, all simulations with $\mathrm{Re}_\mathrm{p} > 50$ were performed using the
unsteady solver, otherwise a steady-state solver was used.

\begin{table}[tb]
	\caption{The information regarding each of the considered geometries and
		computational meshes. The table contains the number of cells per layer
		($N_\mathrm{C}$) and the number of layers ($N_\mathrm{L}$) for each of the
		considered rotation angles $\alpha$.}
	\centering
	\begin{tabular}{lrrrrr}
		\toprule
		$\alpha$                       & $0^\circ$  & $5^\circ$  & $10^\circ$ & $15^\circ$ & $20^\circ$ \\
		$N_\mathrm{L}$                 & 1          & 36         & 18         & 12         & 9          \\
		$N_\mathrm{C}$ ($\times 10^6$) & 1          & 1.46       & 1.33       & 1.22       & 1.46       \\
		\midrule
		$\alpha$                       & $25^\circ$ & $30^\circ$ & $35^\circ$ & $40^\circ$ & $45^\circ$ \\
		$N_\mathrm{L}$                 & 36         & 6          & 36         & 9          & 4          \\
		$N_\mathrm{C}$ ($\times 10^6$) & 1.21       & 1.27       & 1.30       & 1.27       & 1.24       \\
		\midrule
		$\alpha$                       & $50^\circ$ & $55^\circ$ & $60^\circ$ & $65^\circ$ & $70^\circ$ \\
		$N_\mathrm{L}$                 & 18         & 36         & 3          & 36         & 18         \\
		$N_\mathrm{C}$ ($\times 10^6$) & 1.22       & 1.24       & 1.26       & 1.28       & 1.27       \\
		\midrule
		$\alpha$                       & $75^\circ$ & $80^\circ$ & $85^\circ$ & $90^\circ$              \\
		$N_\mathrm{L}$                 & 12         & 9          & 36         & 2                       \\
		$N_\mathrm{C}$ ($\times 10^6$) & 1.27       & 1.26       & 1.25       & 1.26                    \\
		\bottomrule
	\end{tabular}%
	\label{tab:NumberOfCellsAndLayers}
\end{table}

The residual convergence tolerance was set to $1 \times 10^{-13}$ for pressure
and velocity equations in steady-state simulations to ensure proper convergence of
velocity gradients and accurate computation pressure drop over the packed
bed geometry. The maximum number of correctors for mesh non-orthogonality was set to 3.

The unsteady simulations employed a less strict convergence tolerance equal to $1
\times 10^{-8}$. The Courant number was kept below \num{0.8}. The number of PISO 
corrector steps was set to 4, with one additional loop for mesh non-orthogonality.

To force the flow, the pressure gradient momentum source was adopted, with the
value calibrated during runtime, ensuring desired the bulk velocity and Reynolds
number. Second-order accurate spatial discretization schemes using linear
interpolation were used for both convective and diffusive terms. 
For time integration a second-order implicit backward scheme was adopted.

The native geometric–algebraic multigrid (GAMG) solver was predominantly used
for the pressure equation in both steady and unsteady simulations. However, for
rotation angles of $5^\circ$, $10^\circ$, $15^\circ$ and $20^\circ$, at high 
$\mathrm{Re}_\mathrm{p}$, the preconditioned conjugate gradient (PCG) method was used
instead due to convergence difficulties. The PBiCGStab method with the DILU
preconditioner was applied to the velocity equation in all cases.

Each module was meshed separately, resulting in the grids for each geometry
consisting of a number of separate regions, which were connected through a
non-conformal coupling (NCC) mesh interface \citep{greenshields2024}, which
%
%
creates new set of faces at the intersections of overlapping faces on the
coupled patches. These intersection faces become part of the finite volume mesh
on both sides of the interface, ensuring that the mesh is topologically closed.
As a consequence, conservation of mass and momentum is maintained by the
standard finite volume discretisation, without any additional correction. The
coupling is fully conservative and formally second-order accurate.

As with standard unstructured finite volume discretisation such boundary
condition can still result in a decrease of the overall accuracy or convergence
order, resulting from potentially  high non-orthogonality and skewness of
coupled faces, requiring detailed monitoring of the NCC regions. For some of the
considered geometries, we have observed that the use of NCC conditioners led to
the presence of pressure oscillations near the coupled boundaries. As their
presence did not impede convergence and subsequent refinements near the coupled
regions did not contribute to a significant change in averaged results (e.g.,
relative change of friction factor below 2\%), we have deemed the accuracy
offered by NCC acceptable.

Side walls of the module were treated as no-slip boundaries with standard zero
Neumann boundary conditions for pressure. The top and bottom boundaries of each
geometry were also treated using NCC interface, imposing periodic boundary
conditions to replicate an infinitely repeating porous structure.

An unstructured mesh was employed for all cases,
generated using Gmsh \citep{geuzaine2009}, with prism cells near the corners of
smaller void spaces.

\begin{figure}[tb]
	\centering
	\subfigure[]{
		\includegraphics[width=0.4\textwidth]{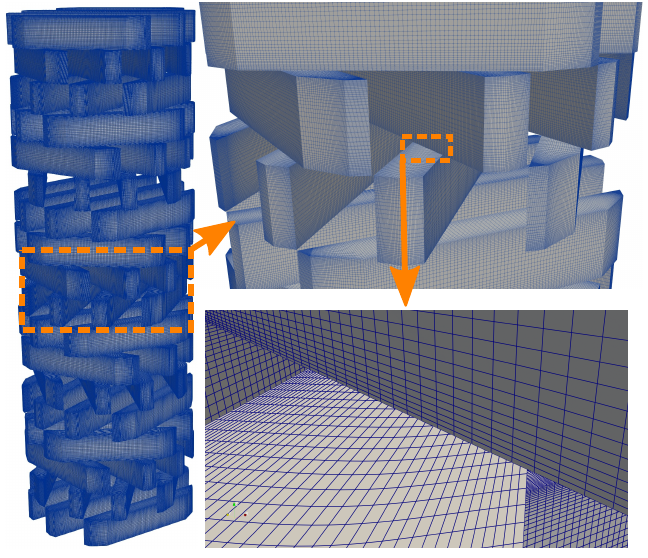}
		\label{fig:angle50Mesh}
	}
	\subfigure[]{
		\raisebox{5mm}{\includegraphics[width=0.4\textwidth]{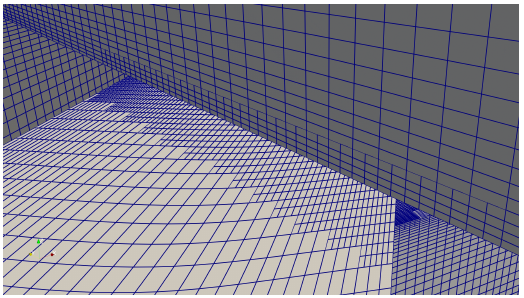}}
		\label{fig:angle50MeshRefined}
	}
	\caption{(a) Schematic representation of the simulation domain for the 
		case with a module angle of $\SI{50}{\degree}$. Part of the domain is
		clipped for clarity. The first zoomed-in view illustrates the orientation
		of the mesh, while the second highlights the mesh structure between
		adjacent layers. (b) Detailed view of the refined mesh at the same angle}
	\label{fig:combinedMeshes}
	
\end{figure}

Sensitivity of the results to the chosen resolution was evaluated using three 
different meshes without refinement (see \cref{app:VerificationMeshSensitivity}
for details). For the steady-state simulations, a medium-resolution mesh
comprising approximately 1 million cells per module was selected. 
Depending on the rotation angle, this configuration produced a maximum
non-orthogonality of $\SI{60}{\degree}$ (with an average of
\SI{9.55}{\degree}) and a maximum skewness of 0.84. As an example,
a mesh with a rotation angle of $\SI{50}{\degree}$ is shown in \cref{fig:angle50Mesh}.

To accurately capture the unsteady flow, for transient simulations, the
meshes were refined near the
overlapping parts of the mesh interfaces between the regions.
The resulting increase in cell count was geometry dependent: for the $\SI{60}{\degree}$
case, this amounted to approximately 20\%, whereas for the $\SI{5}{\degree}$ case, the
increase was around 50\% due to the larger contact surface. After refinement,
the maximum non-orthogonality increased to about  $\SI{70}{\degree}$ (average of
\SI{12.28}{\degree}), and the maximum skewness 1.34. A representative refined
mesh for $50^\circ$ angle case is shown in \cref{fig:angle50MeshRefined}.
\Cref{tab:NumberOfCellsAndLayers} presents the number of cells for each rotated
angle after the refinement.

\begin{table}[tb]
	\centering
	\caption{Normalized simulations times for different Reynolds numbers.
		Here, $T_B = B/\langle w \rangle$ denotes the characteristic time scale,
		$T_\mathrm{i}$ represents the start-up time, and $T_{\mathrm{avg}}$ the average time.}
	\begin{tabular}{c c c}
		\toprule
		$\mathrm{Re}_\mathrm{p}$ & $T_\mathrm{i}/T_B$ & $T_{\mathrm{avg}}/T_B$ \\
		\midrule
		100 & 20 & 10 \\
		150 & 30 & 15 \\
		200 & 40 & 20 \\
		\bottomrule
	\end{tabular}
	\label{tab:NormalizedTimes}
\end{table}
	
For transient simulations, the averaging time was defined such that the
first-order quantities for each Reynolds
number reached a statistically steady-state at randomly selected probe
locations within the simulation domain. The times are listed
in \cref{tab:NormalizedTimes}.

\subsection{Periodic averaging \& data reduction}

\begin{figure} \centering
	\includegraphics[width=0.3\textwidth]{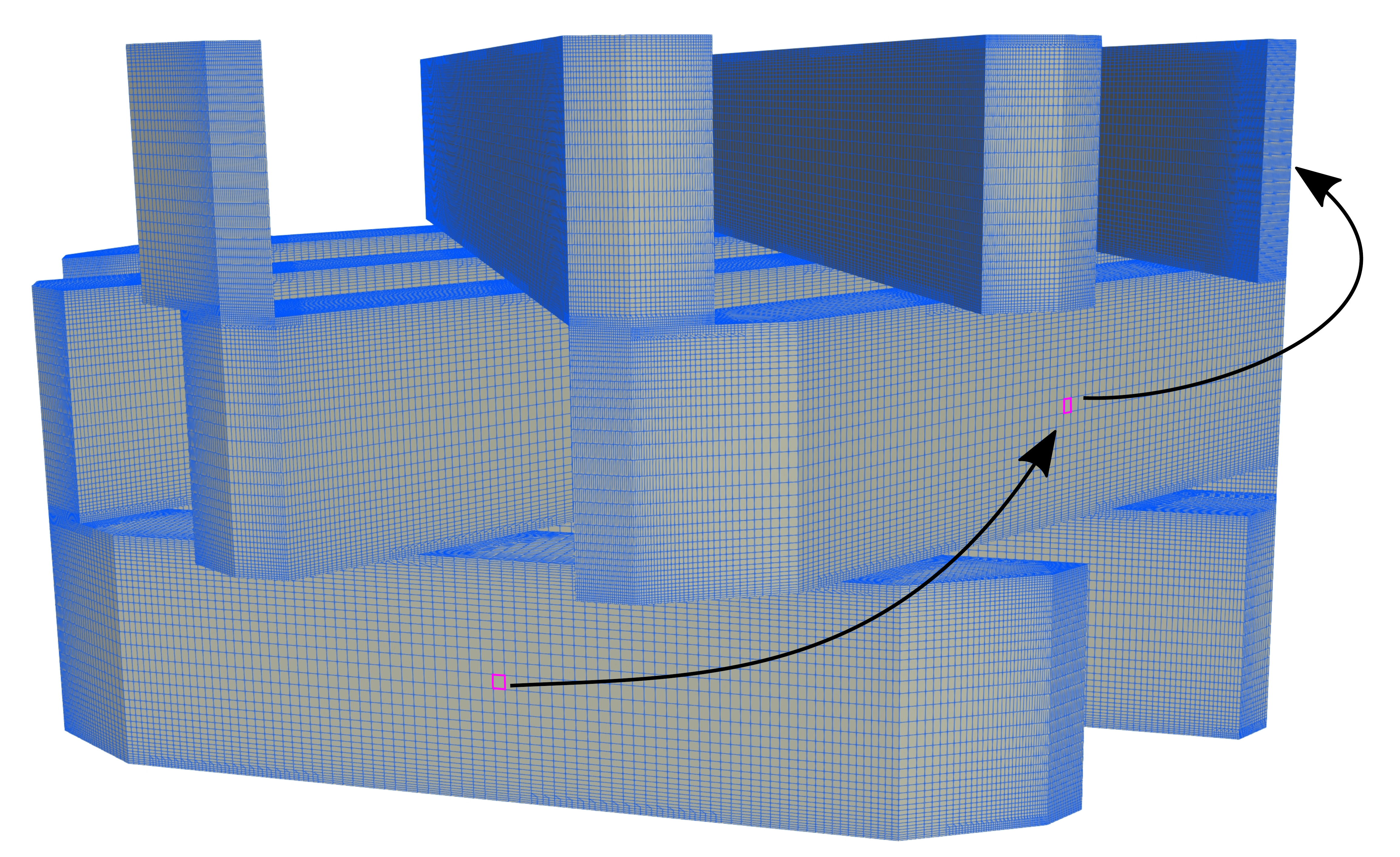} 
	\caption{Computational mesh of the $\SI{60}{\degree}$ 
	simulation domain, representing periodically repeated 
	module used for averaging} \label{fig:dataCompressMesh}
\end{figure}

For each rotation angle $\alpha$, the numerical domain consists of identical
modules, repeated several times as described in \cref{sec:geometry}. The flow
field in the void spaces of each module should be identical when viewed in a
reference frame rotated with the module. This allows to further boost the
statistical convergence of time-averaged data by performing a periodic average,
or averaging the flow field over each module in the considered geometry.

This is achieved by first identifying the corresponding cells and faces, as
illustrated in \cref{fig:dataCompressMesh}. Scalar fields can then be directly
averaged by summing the affiliated field values at these cell/faces and
dividing by the total number of modules. In the case of vector and tensor fields,
before averaging, the field has to be first rotated appropriately for each
module. The average values can be distributed to the cells/faces corresponding
to each other (applying the reverse transformation for the vector and tensor
fields). Both the velocity and pressure fields are averaged this way in this
study.

Finally, since the periodically averaged fields are identical across layers
(without considering the rotation), only one representative module needs to be
considered for analysis and data archiving. The boundary values of the
representative module were calculated by interpolating between adjacent cells.
As an example of the storage efficiency achieved, for  $\alpha=\SI{5}{\degree}$
at $\mathrm{Re}_\mathrm{p} = 200$, the simulation comprising 36 layers originally
required about 15 GB of storage, which was reduced to approximately 135 MB
after averaging. We would like to remark that such method cannot be used for
archiving of snapshots of unsteady velocity or pressure, and we employ it only
for time-averaged data or out of steady-state simulations.
Unless otherwise stated, the results presented in the following
sections are based on periodically averaged mean values.

\section{Results and discussion}\label{sec:results}

\subsection{Validation}

\begin{figure*}[tp]
	\centering
	\subfigure[]{
		\makebox[\textwidth][c]{%
			\includegraphics[width=0.55\textwidth]{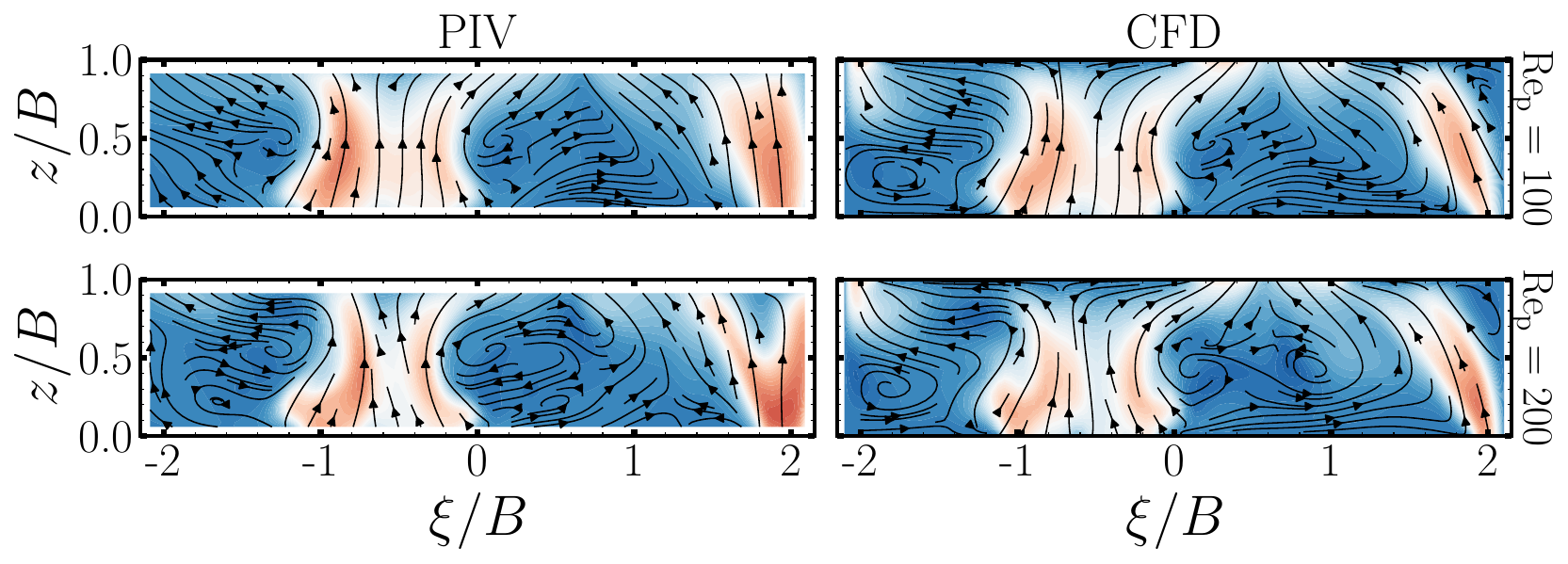}
		}
		\label{fig:_PIV_CFD_contourf_L13_Pos1}
	}\\%
	\subfigure[]{
		\makebox[\textwidth][c]{%
			\includegraphics[width=0.65\textwidth]{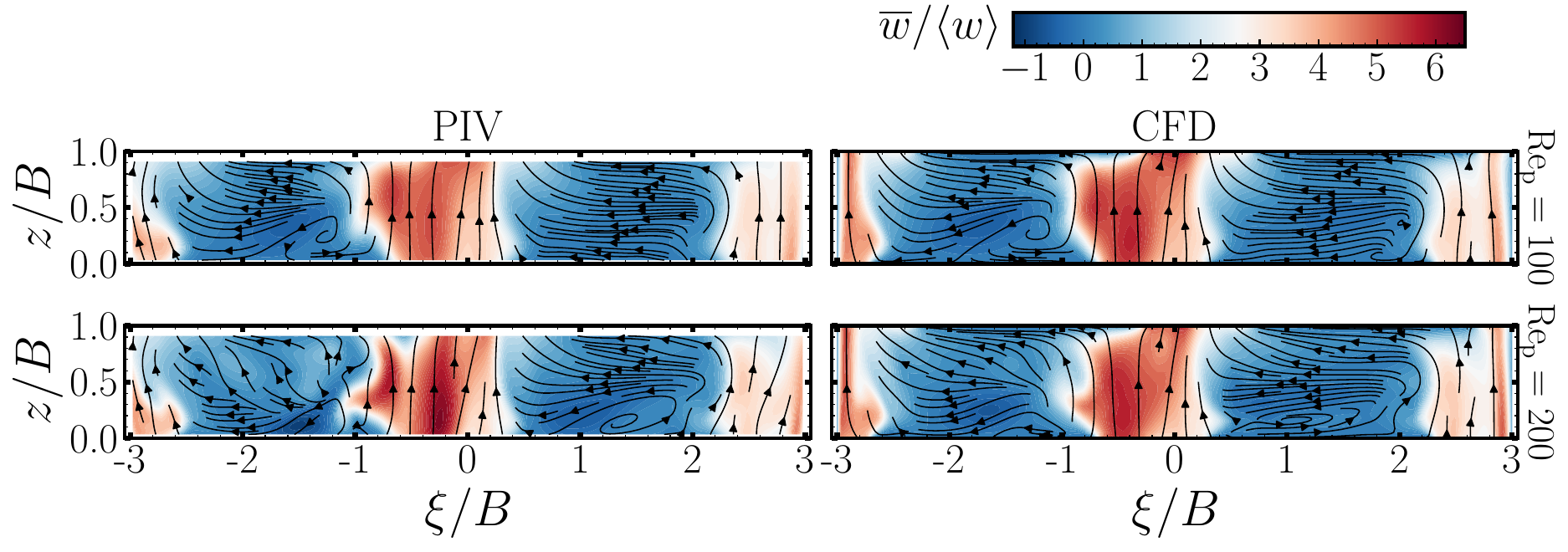}
		}
		\label{fig:_PIV_CFD_contourf_L13_Pos3}
	}
	\caption{Comparison of normalised streamwise velocity fields $\overline{w}/\langle w \rangle$ obtained
		from PIV measurements and CFD simulations at $\alpha = \SI{30}{\degree}$ for
		$\mathrm{Re_p}=100$ and $200$. (a)~Position P1, (b)~position P3.
	}
	\label{fig:_PIV_CFD_contourf_L13}
\end{figure*}

The numerical setup was validated by comparing results for $\alpha=\SI{30}{\degree}$
at $\mathrm{Re}_\mathrm{p}=100, 200$ 
against experiments conducted by
\citet{christin2025}. We compared the mean flow
fields at two planes located between the bars (P1 and P3), as
shown in \cref{fig:ModuleGeometry}. The experimental
configuration consists of 18 modules and an
outlet section, with measurements available at layers
\nth{13}–\nth{17} and above the bed.
Because the velocity at the \nth{13} layer is least affected
by outlet effects, it was selected for comparison with the
fully developed, periodic flow of the simulation.

The flow field at P1 and P3 is compared in
\cref{fig:_PIV_CFD_contourf_L13}, plotted in a $(\xi, z)$ coordinate system 
(horizontal and vertical coordinates, respectively).
The simulation captures the experimental flow features with 
satisfactory accuracy, particularly at the midspan region.
Both datasets indicate that the
flow field behaviour is mainly determined by the geometry of 
the void space, especially by the locations of the inlets 
and outlets. The accelerated fluid near these narrow connections 
between the modules generates several recirculation regions
attached to the bottom and top walls.

While the midspan velocity distributions show strong agreement
in both flow direction and normalized velocity, deviations are
observed at the far left and right sides of the domain. 
These discrepancies are attributed to the experimental setup: 
the RC2 geometry required modification to allow for 
optical access.
For a comprehensive comparative analysis of 
the numerical and experimental results,
the reader is referred to \citet{sadowski2025}.

\begin{figure*}[tb]
	\centering
	\includegraphics[width=0.8\textwidth]{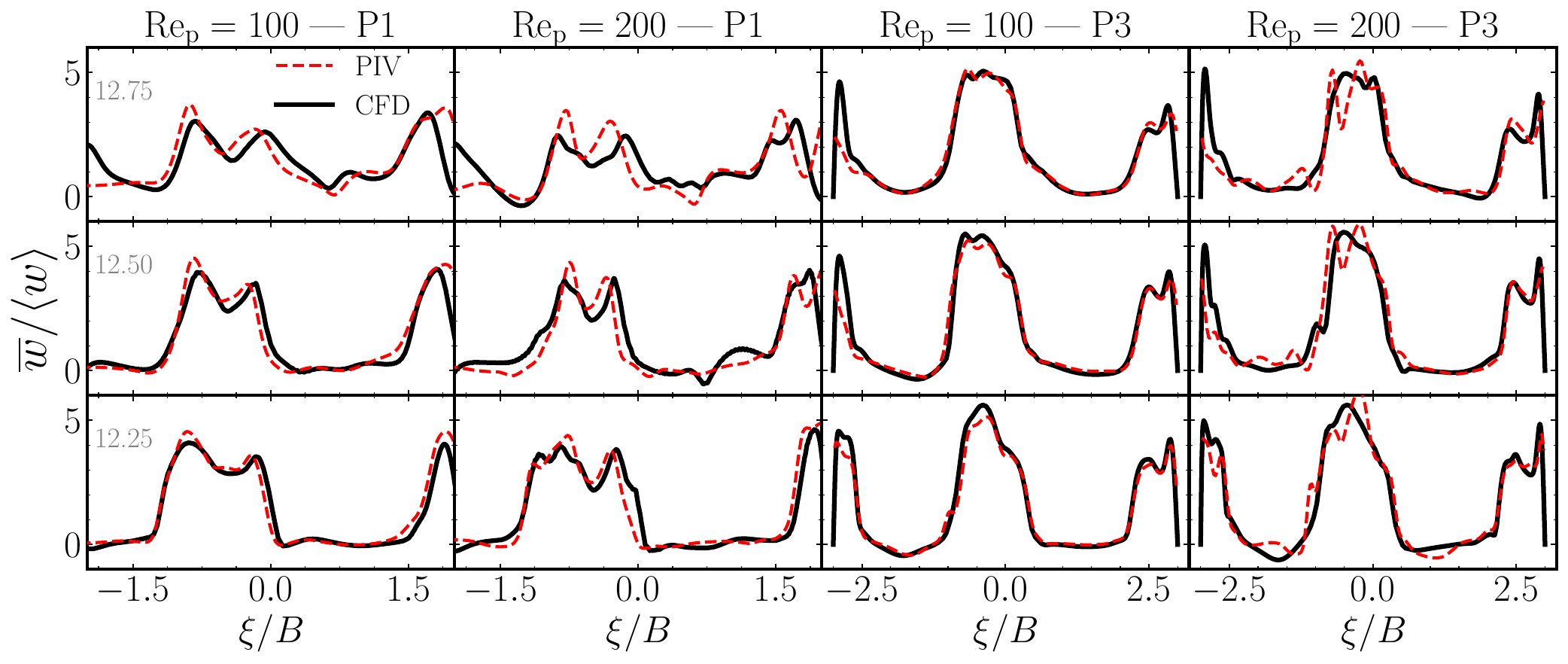} 
	\caption{Normalised spanwise velocity profiles
		($\overline{w}/\langle w \rangle$) from PIV measurements
		and CFD simulations at $\alpha = \SI{30}{\degree}$. Profiles are shown at three vertical
		positions ($z/B = 12.25,\,12.50,\,12.75$)
		for $\mathrm{Re_p}=100$ and $200$.} \label{fig:PIV_CFD_all_row}
\end{figure*}

\Cref{fig:PIV_CFD_all_row} presents the spanwise
velocity profiles normalized by the intrinsic velocity
at three $z$-positions of the \nth{13} layer for
$\mathrm{Re}_\mathrm{p}=100$ and $200$. 

At $\mathrm{Re}_\mathrm{p}=100$, the simulations reproduce the
characteristic flow features with high fidelity, 
capturing both the location and magnitude of the peak velocities.
This quantitative agreement is confirmed by the root mean square
deviation (RMSD) analysis, which yielded its lowest value (0.333)
at this Reynolds number at location P3.
However, three notable deviations are observed. First, 
minor discrepancies appear near the lateral boundaries, 
particularly at $z=12.75\ \mathrm{mm}$. Second, a boundary-layer jet is
clearly resolved at P3 in the CFD data but is essentially 
absent in the PIV measurements; both are likely due to the 
geometric differences in the experimental setup discussed 
previously. Third, at the far left and right sides,
P3 exhibits a jet-like flow while P1 does not
(a trend also observed at $\mathrm{Re}_\mathrm{p}=200$).
This is attributable to the location of P3, 
which is closer to the central axis, at this orientation.

At $\mathrm{Re}_\mathrm{p}=200$, the flow is unsteady,
increased inertial effects lead to
significantly stronger vortices and the 
emergence of distinct local velocity fluctuations. 
Reflecting this increased physical complexity,
the RMSD at P1 is 1.009. Although this represents
the maximum deviation across the simulated cases,
it remains within acceptable limits considering
the transient nature of the flow, and the overall
agreement between the numerical and experimental 
results remains robust.

\subsection{Overview of the flow fields}

\begin{figure*}[t]
	\centering
	\subfigure[]{
		\includegraphics[width=0.30\textwidth]{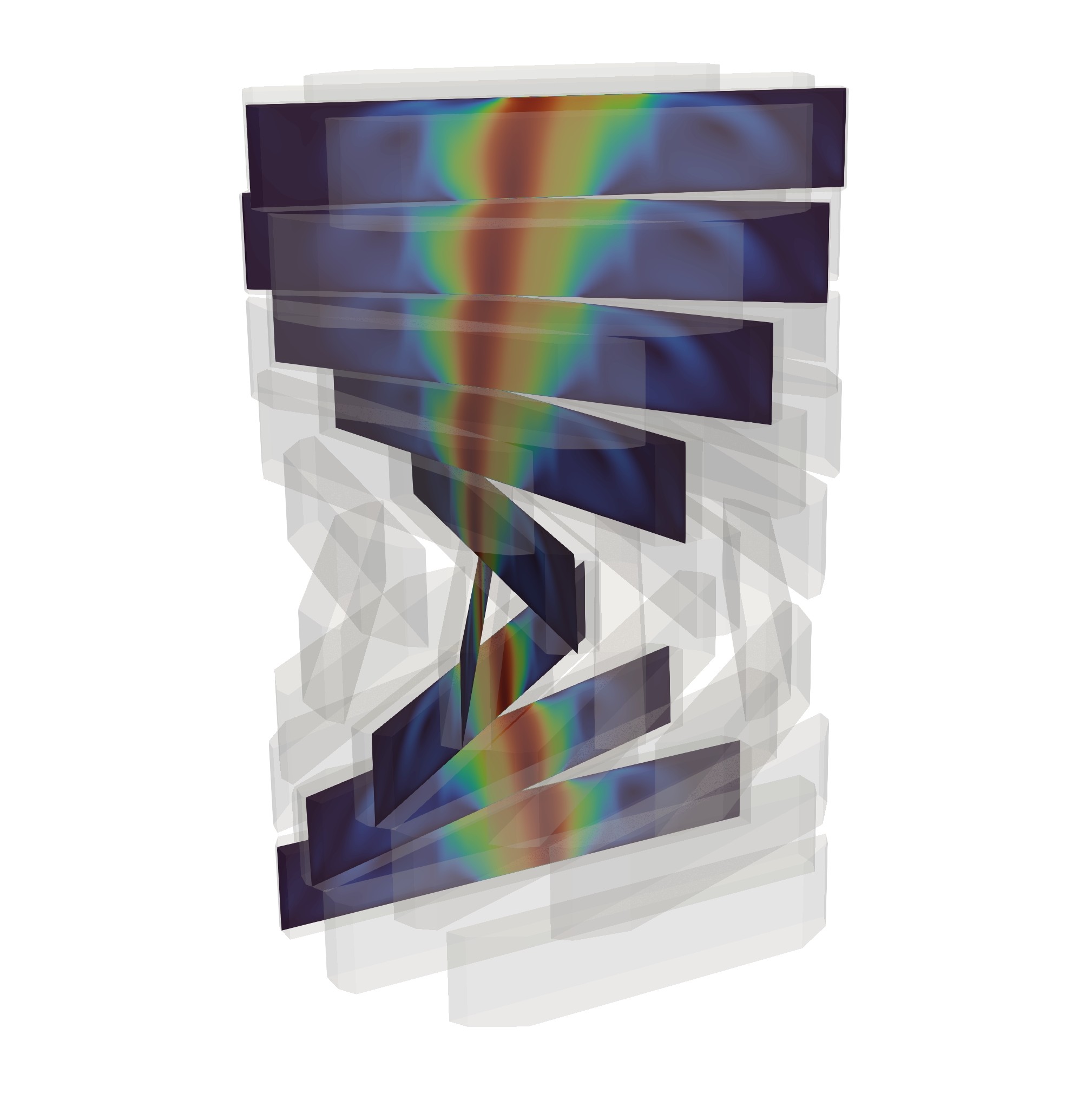} \label{fig:Re200_a20}}
	 \raisebox{0mm}{\subfigure[]{
			\includegraphics[width=0.35\textwidth]{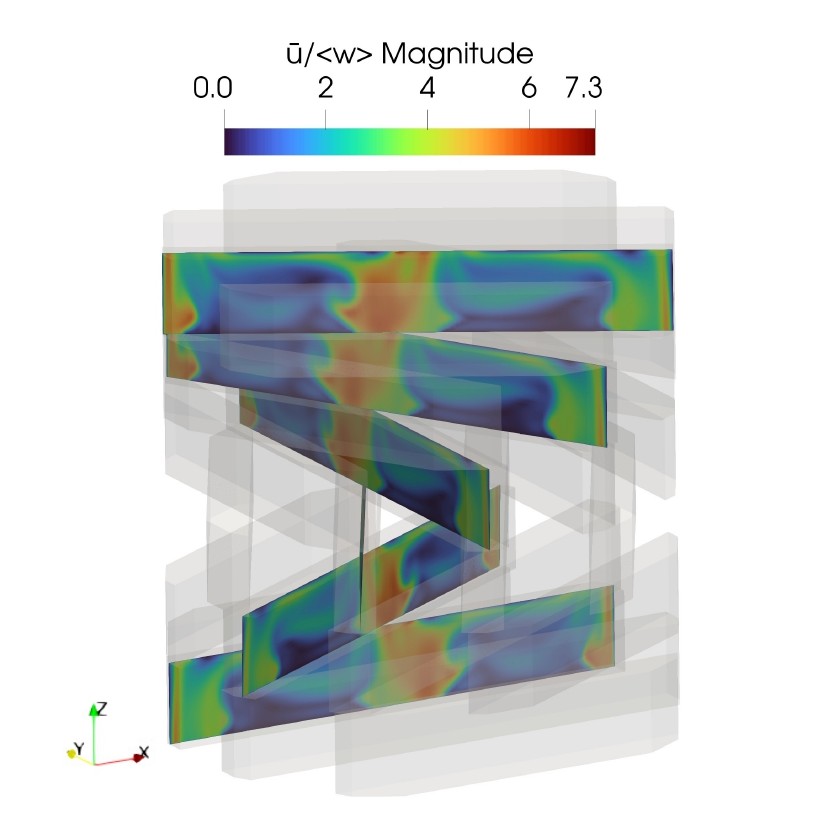}}
		\label{fig:Re200_a30}%
	}
	\caption{%
		Normalized mean velocity magnitude without periodic averaging at $\mathrm{Re}_\mathrm{p} = 200$ 
		for (a) $\SI{20}{\degree}$ and (b) $\SI{30}{\degree}$ geometries, visualized 
		on the plane at position P3 (see \cref{fig:ModuleGeometry}). The flow direction is along the $+z$ axis.
	} \label{fig:Re200_a20_a30}
\end{figure*}

We provide an overview of the flow field under different geometric 
configurations and flow conditions. 
\Cref{fig:Re200_a20_a30} presents the mean
velocity field for two different angles ($\alpha=\SI{20}{\degree},
\SI{30}{\degree}$) at the highest Reynolds number ($\mathrm{Re}_\mathrm{p} =
200$), on the slices through the void spaces in each module located at the
position P3 (see \cref{fig:ModuleGeometry}).
The strong similarities observed between successive layers highlight the periodic 
nature of the flow, thereby justifying the use of periodically averaged
quantities in the following sections.

For the rotation angle of $\SI{20}{\degree}$, visualized in
\cref{fig:Re200_a20}, the areas of high velocity field are confined to the
central section of the void spaces. Although the bars at this angle are still
partially aligned, the geometry already exhibits characteristics of the
\emph{lattice-like} regime: the void spaces show increased lateral connectivity
compared to smaller angles ($\alpha \leq \SI{10}{\degree}$). Each void space retains
a degree of streamwise connectivity (closely aligned inlet and outlet), inducing
preferential transport of momentum along the centre. Moreover, a weak
recirculating flow can be observed in the cavity-like regions of the void
spaces located at the sides of the strong core flow. This is consistent with
the observations of \citet{Zhang2025}, that in the inertial regime of laminar
flow, recirculation zones emerge near the walls when the flow passes through
pore-throat structures.
\Cref{fig:Re200_a30} shows the mean velocity field at the same location in each layer
for the $\SI{30}{\degree}$ geometry.
At this rotation angle, the layers become multiply-connected with inlets and
outlets staggered due to the rotation of the base geometry. We can refer to such geometry as
\emph{lattice-like}, and in such cases, the interactions between successive
layers lead to a more heterogeneous velocity distribution than in the
channel-like geometries ($\alpha \leq \SI{10}{\degree}$). In the $\alpha = \SI{30}{\degree}$ case, 
the positioning of the inlets and outlets results in pronounced wall-jets
(on the order of $4$–$5\langle w\rangle$) and large recirculation regions 
characterized by high velocities, leading to a highly non-uniform distribution
of velocity within the void spaces. 
For a more in-depth study of the flow in this geometry the reader is referred to our previous work \citep{sadowski2025}.

\begin{figure*}[tb]
	\centering
	\subfigure[]{
		\includegraphics[width=0.32\textwidth]{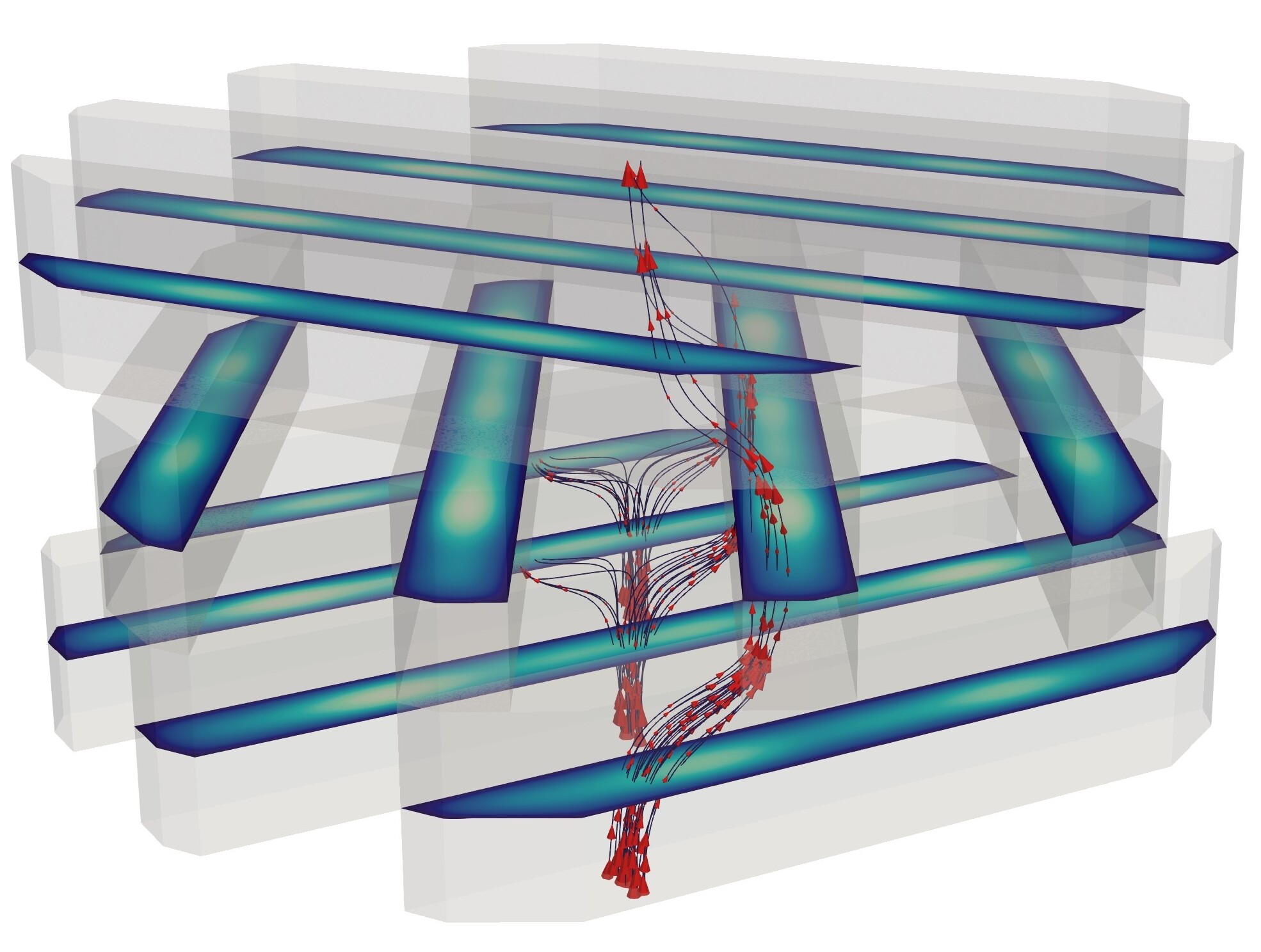} \label{fig:Re1_a60}}
	\subfigure[]{
		\includegraphics[width=0.32\textwidth]{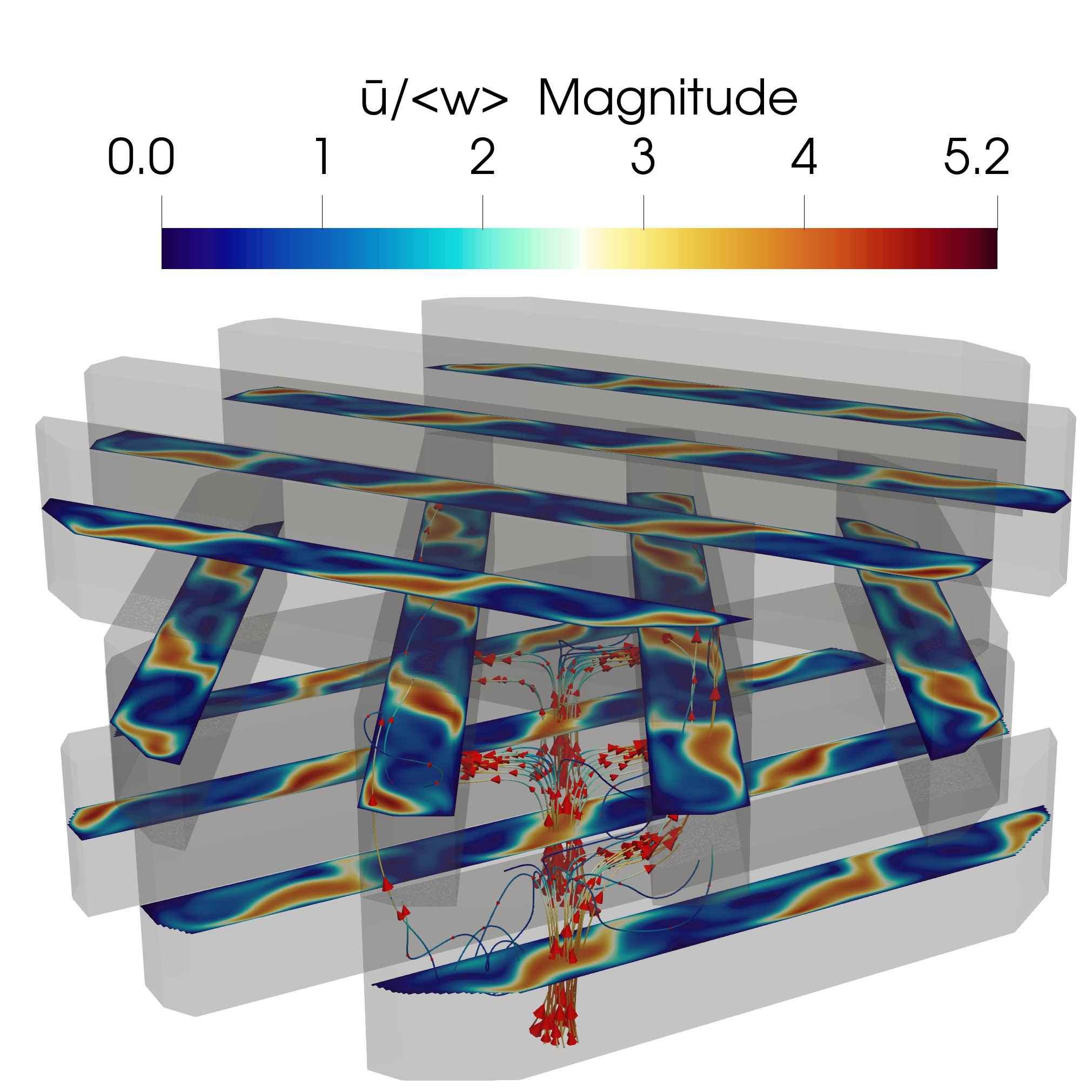}\label{fig:Re200_a60}}
	\caption{Normalized mean velocity magnitude without periodic averaging in a packed bed with a rotation angle
	of $\SI{60}{\degree}$. Slices are taken from the middle of each module: 
	(a) at $\mathrm{Re}_\mathrm{p} = 1$, (b) at $\mathrm{Re}_\mathrm{p} = 200$. } \label{fig:Re1_200_streamlines}
\end{figure*}

After examining the influence of the rotation angle on the flow structure, the
effect of the Reynolds number on the flow regime was considered.
\Cref{fig:Re1_200_streamlines} shows the normalized mean velocity magnitude
for a geometry with $\alpha = \SI{60}{\degree}$ at two different Reynolds numbers.
The slices shown in \cref{fig:Re1_200_streamlines} are extracted from the
mid-planes of three successive modules, located at $z = 5$, $15$, and $\SI{25}{mm}$, respectively.

At $\mathrm{Re}_\mathrm{p} = 1$ (\cref{fig:Re1_a60}), the flow is fully laminar and stationary with 
the pressure drop primarily governed by Darcy's law. 
The fluid passes smoothly through the void spaces between layers, where the normalized 
velocity magnitude ranges from 0 to 2.5.
The streamlines follow
smooth, slightly curved
trajectories directed upward through the bed.
The velocity distribution within the void space exhibits
a low-speed, upward-directed flow that accelerates toward the center.

At $\mathrm{Re}_\mathrm{p} = 200$ (\cref{fig:Re200_a60}), 
the flow is in the inertia-dominated regime.
In contrast to $\mathrm{Re}_\mathrm{p} = 1$ (\cref{fig:Re1_a60}), 
the streamlines here exhibit significant tortuosity.
This complexity arises because the increased inertial
forces prevent the fluid from following the curvature of the 
solid bars, leading to the formation of low-velocity wake
regions and distinct recirculation zones within the void spaces.
The flow develops strong transverse velocity components,
which drive significant deviations from the primary streamwise direction.

\subsection{Friction factor}

\begin{figure}[tb] \centering
	\includegraphics[width=0.85\linewidth]{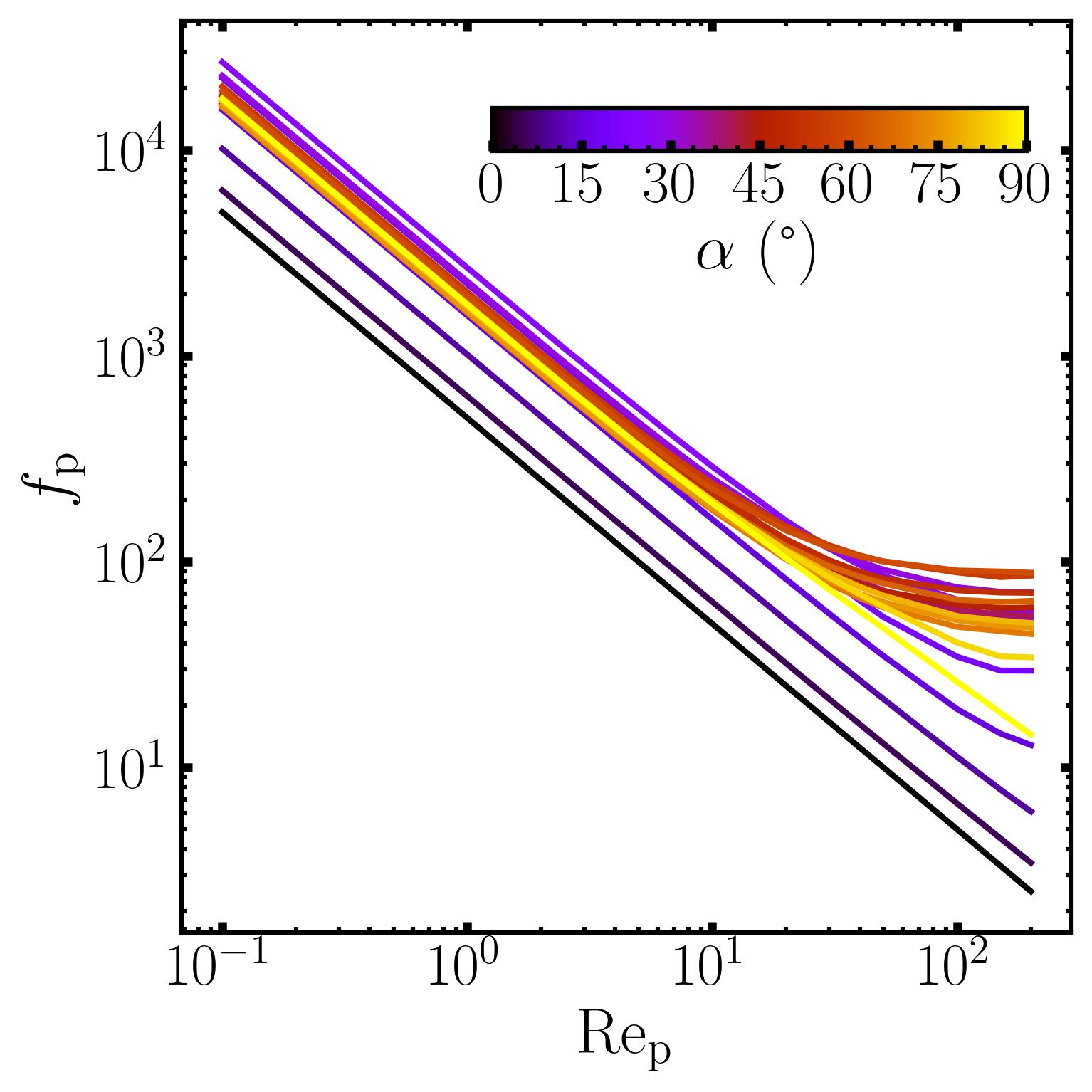} 
	\caption{Friction factor ($f_{\mathrm{p}}$) as a function of $\mathrm{Re}_{\mathrm{p}}$ for different rotation angles}
	\label{fig:friction_factor_reynold_num}
\end{figure}

\begin{figure}[tb] \centering
	\includegraphics[width=0.85\linewidth]{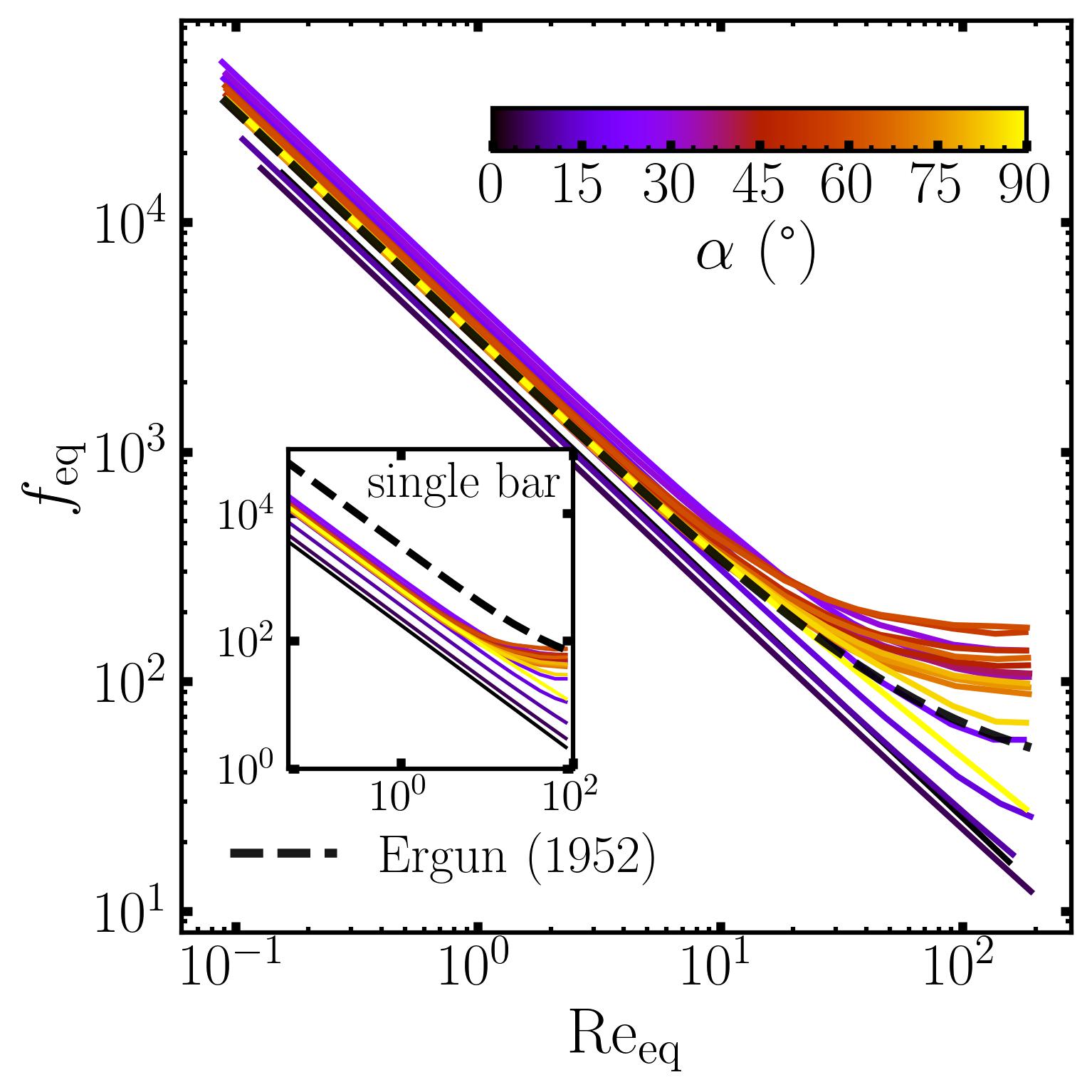} 
	\caption{Friction factor ($f_{\mathrm{eq}}$) as a function of
		$\mathrm{Re}_{\mathrm{eq}}$ for different rotation angles.
		The dotted black line shows the Ergun correlation (\cref{eq:friction_factor_Erg}).
		Inset: same data computed with the single-bar diameter~$d_{\mathrm{sb}}$.}
	\label{fig:friction_factor_reynold_num_deq}
\end{figure}

\Cref{fig:friction_factor_reynold_num} shows the relation of the  friction
factor as a function of Reynolds number for various angles. Each line represents
the friction factor calculated from \cref{eq:friction_factor}. The color
gradient indicates the rotation angle of each geometry from $\SI{0}{\degree}$ to
$\SI{90}{\degree}$.

In the viscous regime, all curves follow a hyperbolic trend with respect to
$\mathrm{Re}_{\mathrm{p}}$, which appears as a straight  line in the log--log
scale. As the angle increases from $\alpha = \SI{0}{\degree}$ to
$\SI{25}{\degree}$, the friction factor increases and reaches its maximum at
$\alpha = \SI{25}{\degree}$. Beyond this angle, the friction factor decreases
gradually, and further changes in rotation angle no longer introduce a large
variation in $f_{\mathrm{p}}$.

As the Reynolds number increases, deviations from the linear trend become
apparent, signaling the onset of inertial effects and the transition toward the
non-linear (Forchheimer) regime, for larger angles. However, the configuration
at $\alpha = \SI{90}{\degree}$ exhibits a much slower transition to the inertial
regime (see \cref{sec:transition}) with $f_{\mathrm{p}}$ resembling a
straight line.
Additionally, the angle yielding the maximum
friction factor shifts with the Reynolds number. For instance, at
$\mathrm{Re}_{\mathrm{p}} = 200$,  the maximum friction factor occurs at
$\alpha = \SI{60}{\degree}$, in contrast to the peak at
$\alpha = \SI{25}{\degree}$ observed in the viscous regime.

These values of the drag can be traced to the properties of each individual
geometry. The specific alignment of the square bars at
$\alpha = \SI{25}{\degree}$ creates a severe constriction of the flow path.
These narrowing void spaces increase the local shear stress, resulting  in the
maximum friction factor observed in the viscous regime.  At
$\alpha = \SI{60}{\degree}$, the flow shows behaviour similar to
contraction–expansion sequences in ducts. As the Reynolds number increases, the
acceleration into void spaces creates local jets, leading to flow separation and
recirculation downstream of these constrictions.  Consequently, form drag
becomes the dominant loss mechanism, resulting in the highest observed friction
factor. 
In the case of $\alpha=\SI{90}{\degree}$, the perfect orthogonality between
modules allows the flow to develop with minimal obstruction, resulting in
reduced flow resistance. Despite this low friction factor, the geometric
parameters of the \SI{90}{\degree} configuration ($D/d_{\mathrm{eq}} = 3.23$,
$a_v = 0.193~\mathrm{mm^{-1}}$) place it within the lattice-like group rather
than the channel-like regime observed at $\alpha \leq \SI{10}{\degree}$.



The accuracy of the Ergun depends strongly on the choice of
characteristic length scale.
\Cref{fig:friction_factor_reynold_num_deq} presents the friction factor
computed with the module-equivalent diameter~$d_{\mathrm{eq}}$. In the
viscous regime, the curves for the lattice-like geometries
($\alpha \geq \SI{15}{\degree}$) collapse onto the Ergun correlation. This is
attributed to the fact that~$d_{\mathrm{eq}}$ incorporates the
angle-dependent wetted surface area through~$a_v$
(\cref{eq:d_eq,eq:sauter-mean-diam}), unlike~$d_{\mathrm{sb}}$, which
remains constant for all rotation angles.
With increasing Reynolds number, the curves progressively
diverge: at intermediate angles (notably $\alpha \approx \SI{60}{\degree}$) the
friction factor exceeds the Ergun prediction, whereas the channel-like
configurations ($\alpha \leq \SI{10}{\degree}$) and $\alpha = \SI{90}{\degree}$ remain
below it.

When the constant single-bar diameter~$d_{\mathrm{sb}}$ is employed instead
(inset of \cref{fig:friction_factor_reynold_num_deq}), the Ergun correlation
overestimates the friction factor systematically across all angles and
Reynolds numbers. Furthermore, the spread among the individual curves is
considerably larger, since~$d_{\mathrm{sb}}$ does not vary with~$\alpha$ and
is therefore unable to capture the differences in wetted surface area between
configurations. These observations indicate that~$d_{\mathrm{eq}}$ is the
more suitable length scale for correlating the friction factor in the present
geometry.


\subsection{Determining permeability and forchheimer coefficient}

The determination of the $\mathrm{Da}$ and $C_\mathrm{F}$ numbers was carried
out in two steps using a least-squares fitting optimization procedure of
\cref{eq:friction_factor_nondimensional} to the gathered friction factor data.
First, the Darcy number was obtained for $\mathrm{Re} \le 1$ by neglecting
inertial effects (i.e.\ assuming  $C_\mathrm{F} = 0$). In the second step, the
$C_\mathrm{F}$ was evaluated by fitting the
\cref{eq:friction_factor_nondimensional} over the entire Reynolds number range
while keeping the previously determined Darcy number fixed.

\begin{figure}[tb]
  \centering
  \includegraphics[width=0.4\textwidth]
	{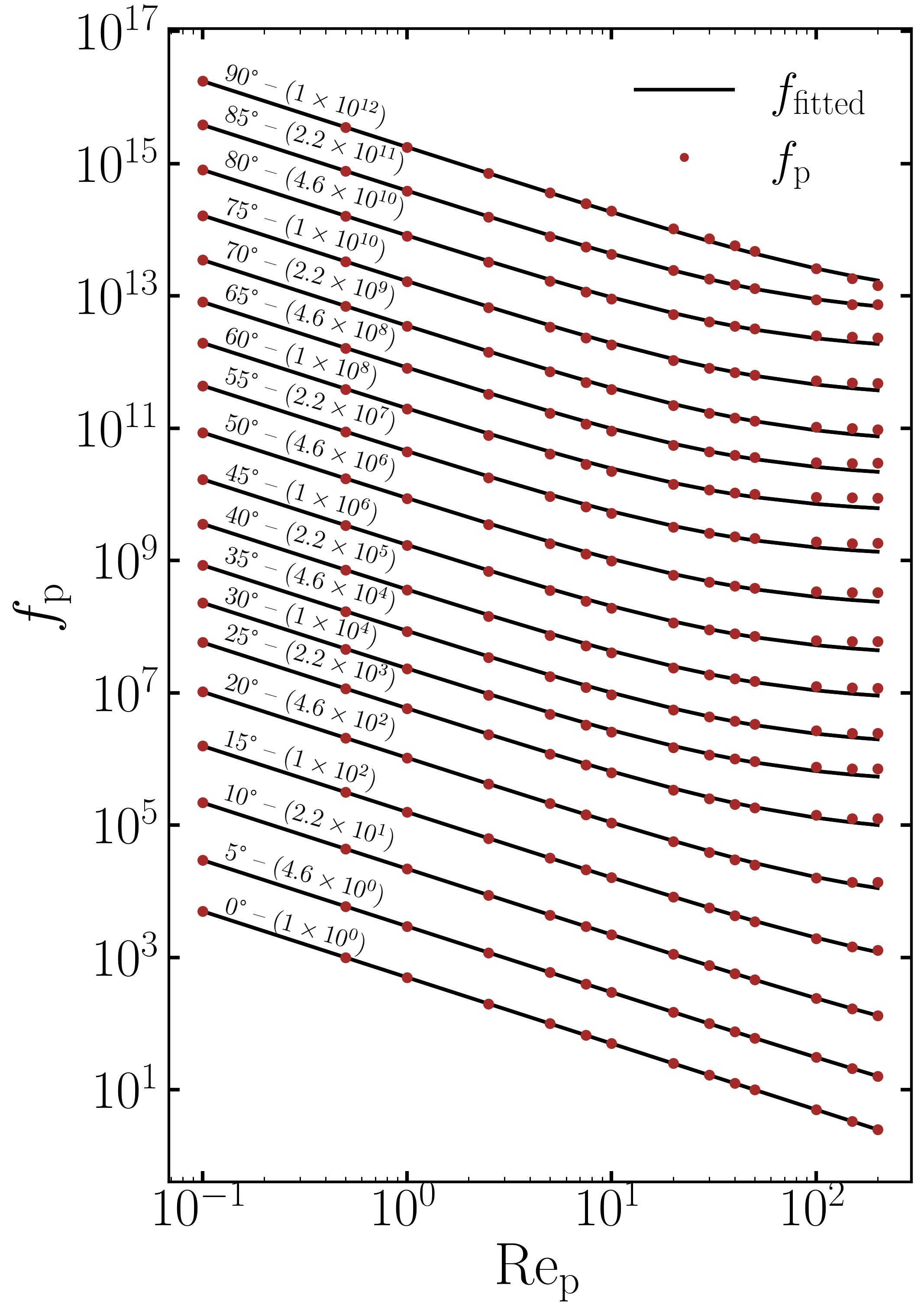}
	\caption{Comparison between the computed friction factor
	($f_\mathrm{p}$) and the fitted friction factor
	($f_\mathrm{fitted}$) as a function of $\mathrm{Re}_\mathrm{p}$
	for different angles. Scaling factors are applied to separate the
	curves vertically for clarity.} \label{fig:friction_factor_fitting}
\end{figure}

\Cref{fig:friction_factor_fitting} presents a comparison between  the computed
friction factor ($f_\mathrm{p}$), obtained from \cref{eq:friction_factor}, and
the fitted non-dimensional friction factor ($f_\mathrm{fitted}$). The latter was
evaluated using  \cref{eq:friction_factor_nondimensional}, incorporating the
calculated Darcy number and Forchheimer coefficient.  As shown in
\cref{fig:friction_factor_fitting}, the fitted curves closely follow the
computed data over the selected range of Reynolds numbers. The agreement is
strong in the viscous regime, while slight deviations appear at higher Reynolds
numbers, suggesting that additional simulations at higher
$\mathrm{Re}_\mathrm{p}$ may be required to better capture the inertial effects.

\begin{figure}[tb]
	\centering
	\subfigure[]{
		\includegraphics[width=0.32\textwidth]{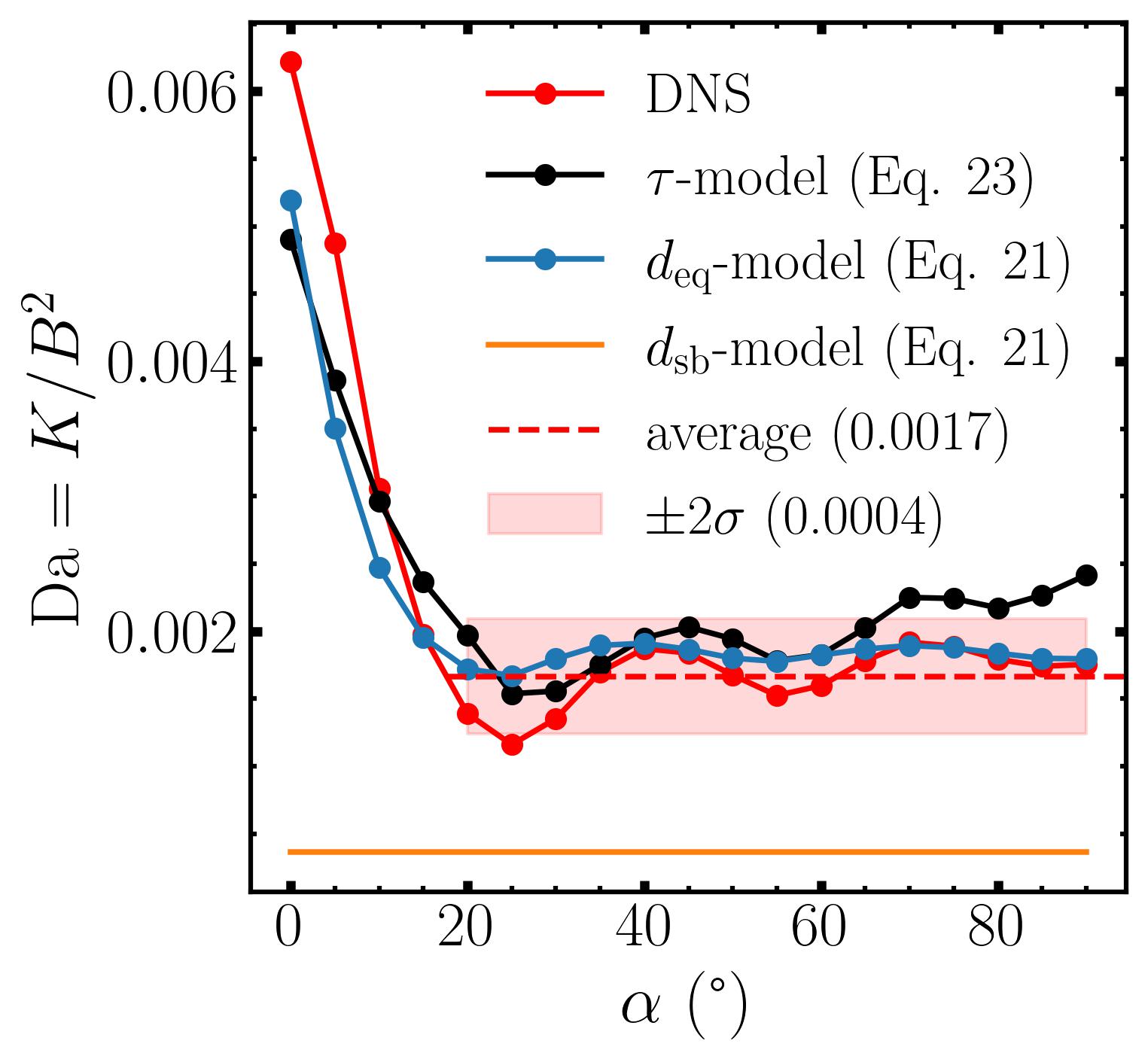} \label{fig:Darcy_angle}}
	\subfigure[]{
		\includegraphics[width=0.32\textwidth]{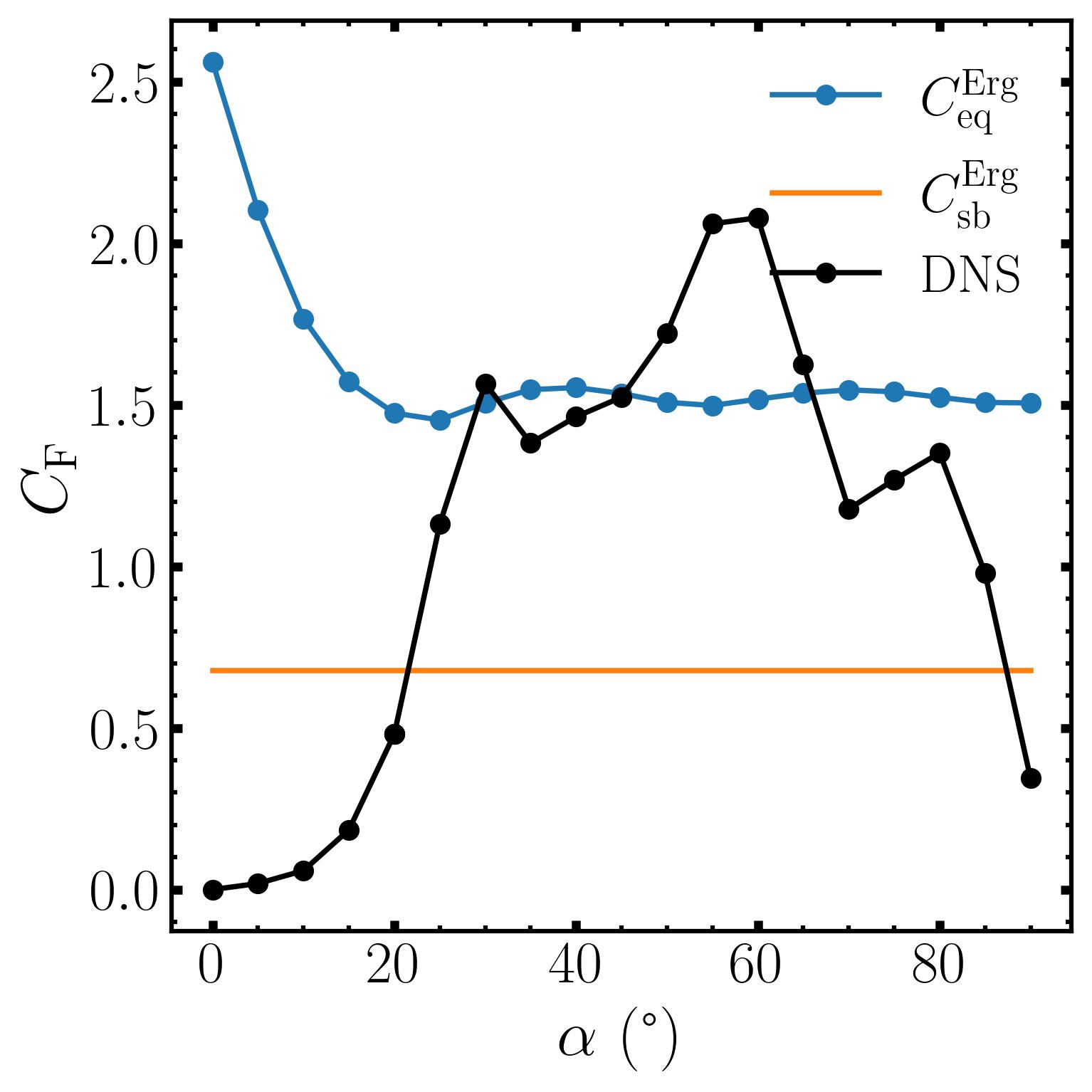}\label{fig:Cf_number}}
	\caption{Darcy--Forchheimer model parameters as a function of the rotation angle~$\alpha$. 
		(a)~Darcy number $\mathrm{Da} = K/B^2$, the tortuosity-based model~($K_\tau$) and                                                                                                               
		Blake--Kozeny correlations using~$d_{\mathrm{eq}}$ and~$d_{\mathrm{sb}}$.
		The dashed line and shaded band indicate the DNS average ($0.0017$) and
		$\pm 2\sigma$ ($0.0004$), respectively.
		(b)~Forchheimer coefficient~$C_\mathrm{F}$ obtained from DNS fitting
		compared with values computed using~$d_{\mathrm{sb}}$ and~$d_{\mathrm{eq}}$. } \label{fig:Da_Cf_angle}
\end{figure}

The computed Darcy number, plotted in \cref{fig:Darcy_angle}, shows a strong
dependence on the rotation angle, demonstrating two distinct behaviours depending
on the geometry type. For $\alpha \le \SI{10}{\degree}$, the gaps in
consecutive modules are aligned in a \emph{channel-like} form, which promotes a
less perturbed flow and results in high permeability. As the angle increases, the
connections between consecutive modules become more obstructed, leading to a sharp
decrease in permeability with $\alpha$, observable also as a previously described trend
in the friction factor in the viscous regime.

For $\alpha \geq \SI{15}{\degree}$, the interstitial space becomes multiply
connected, leading to a stabilization in the value of $\mathrm{Da}$, with a
sinusoidal-like oscillations appearing as different configurations of
connections between consecutive layers promote different values of permeability.
The red dashed line indicates the mean Darcy number ($\mathrm{Da} = 0.0017$) for
$\alpha \geq \SI{15}{\degree}$ geometries, and the magnitude of the oscillations is
characterised by a standard deviation of $\sigma=0.0002$ (a coefficient of
variation of approximately $\SI{12}{\percent}$). The minimum Darcy number was
obtained at $\alpha = \SI{25}{\degree}$, in agreement with $f_{\mathrm{p}}$
reaching maximum at the same angle in the viscous regime.

The configurations at $\alpha = \SI{15}{\degree}$,
$\SI{20}{\degree}$ and $\SI{90}{\degree}$ are classified as lattice-like
based on their bed-to-particle diameter ratio ($D/d_{\mathrm{eq}} > 3$).
This classification is confirmed by their Darcy numbers
($\mathrm{Da} = 0.0020$, $0.0019$ and $0.0018$, respectively), which fall
within the range of the lattice-like group
($\mathrm{Da} = 0.0017 \pm 0.0004$). For the
$\alpha = \SI{90}{\degree}$ configuration in particular,
$D/d_{\mathrm{eq}} = 3.23$ and
$a_v = 0.193~\mathrm{mm^{-1}}$ are consistent with the lattice-like
geometries, in contrast to the channel-like configurations at
$\alpha \leq \SI{10}{\degree}$, where $D/d_{\mathrm{eq}} < 3$ and
$a_v < 0.16~\mathrm{mm^{-1}}$. This confirms that the bed-to-particle
diameter ratio ($D/d_{\mathrm{eq}} \approx 3$ as boundary) and the
hydraulic tortuosity ($\tau \approx 1.1$, see
\cref{fig:tort_eff_Re_angle}) provide a reliable basis for the
classification.

The Forchheimer coefficient~$C_\mathrm{F}$, shown in \cref{fig:Cf_number},
exhibits a trend that parallels the Darcy number at small rotation angles:
channel-like configurations ($\alpha \leq \SI{10}{\degree}$) yield the lowest
values of~$C_\mathrm{F}$. This is not a direct consequence of the high
permeability at these angles, but rather reflects a common geometric origin:
aligned void spaces suppress flow separation and recirculation, thereby
reducing both viscous resistance (high~$\mathrm{Da}$) and inertial losses
(low~$C_\mathrm{F}$).

For $\alpha = \SI{90}{\degree}$, the low~$C_\mathrm{F}$ is consistent with the
delayed transition to the inertial regime observed for this configuration
(see \cref{sec:transition}), which suggests that the Forchheimer term is
not fully developed within the present Reynolds number range.

At intermediate angles ($\SI{30}{\degree}$--$\SI{70}{\degree}$), $C_\mathrm{F}$ exhibits
considerably stronger variations with~$\alpha$ than at the extremes. In this
range, the void-space topology changes significantly with each increment in
rotation angle, producing markedly different levels of flow acceleration,
separation and recirculation.

\subsection{Modelling permeability and forchheimer coefficient}


\begin{figure*}[tb] \centering	\includegraphics[width=0.8\textwidth]{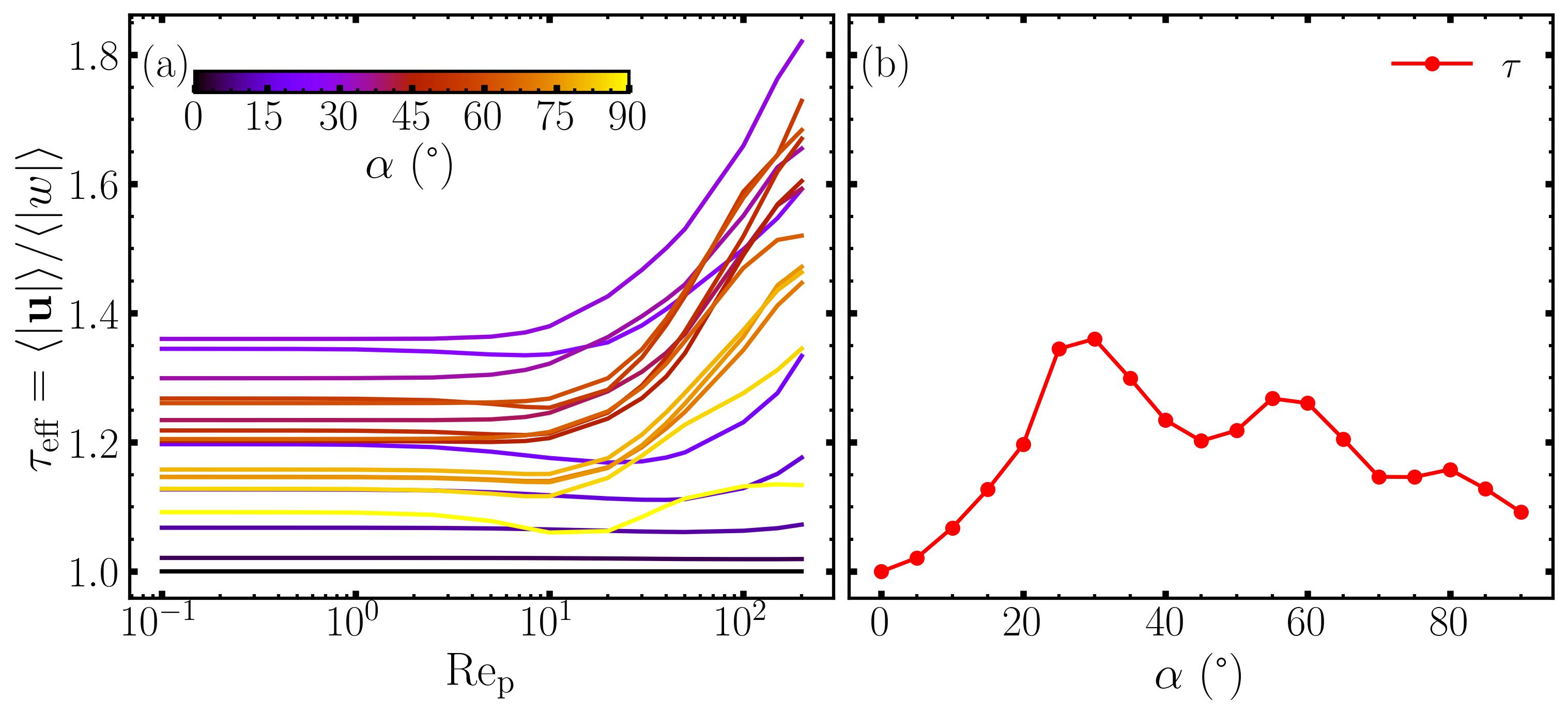}
	\caption{(a) $\tau_{\mathrm{eff}}$ (\cref{eq:tortuosity-eff})
		as a function of~$\mathrm{Re}_\mathrm{p}$ for different rotation
		angles.
		(b) Hydraulic tortuosity~$\tau$ (\cref{eq:tortuosity}) as a
		function of the rotation angle~$\alpha$.} \label{fig:tort_eff_Re_angle}
\end{figure*}

\Cref{fig:tort_eff_Re_angle}(a) shows the upper bound for the hydraulic
tortuosity~$\tau_{\mathrm{eff}}$, computed from \cref{eq:tortuosity-eff}, as a
function of the Reynolds number for all rotation angles. In the viscous regime,
$\tau_{\mathrm{eff}}$ remains approximately constant for each angle, consistent
with the Darcy flow assumption where the flow topology is independent of the
flow rate. As the Reynolds number increases, inertial effects alter the flow
paths and $\tau_{\mathrm{eff}}$ increases for most angles. The channel-like
configurations ($\alpha \leq \SI{10}{\degree}$) exhibit the lowest values
($\tau_{\mathrm{eff}} \approx 1$), reflecting nearly straight flow paths.
The $\alpha = \SI{90}{\degree}$ configuration also shows a low tortuosity
($\tau_{\mathrm{eff}} \approx 1.09$), placing it at the lower bound of
the lattice-like group.

\Cref{fig:tort_eff_Re_angle}(b) presents the hydraulic tortuosity~$\tau$
(\cref{eq:tortuosity}) as a function of the rotation angle. For
$\alpha \leq \SI{15}{\degree}$, $\tau$ is close to unity, indicating that
the aligned bar arrangement allows the flow to pass with minimal deviation.
A sharp increase occurs between $\alpha = \SI{10}{\degree}$ and
$\alpha = \SI{25}{\degree}$, coinciding with the transition from channel-like to
lattice-like geometries. For $\alpha \geq \SI{15}{\degree}$, $\tau$
oscillates between approximately $1.1$ and $1.35$, with a peak near
$\alpha = \SI{25}{\degree}$. This is consistent with the friction factor results
(\cref{fig:friction_factor_reynold_num}), where the maximum drag in the
viscous regime was also observed at~$\alpha = \SI{25}{\degree}$.


\Cref{fig:Darcy_angle} compares the DNS permeability with three model
predictions. The Blake--Kozeny estimate based on the module-equivalent
diameter, $K_\mathrm{eq}$ (\cref{eq:K-both}), provides the
closest agreement with the DNS data for the lattice-like geometries
($\alpha \geq \SI{15}{\degree}$), with a mean absolute percentage error
(MAPE) of $12.3\%$. Because~$d_\mathrm{eq}$ incorporates the
angle-dependent wetted surface area through~$a_v$
(\cref{eq:sauter-mean-diam,eq:equiv-diam}), it captures the dominant
geometric effect governing permeability variations in these configurations.

The tortuosity-based estimate~$K_\tau$ (\cref{eq:K-tau}),
which combines the hydraulic tortuosity~$\tau$ with the specific surface
area~$a_v$ and a fitted Kozeny constant $c_{\mathrm{KC}} = 3.1$,
reproduces the angle-dependent trend with a MAPE of $18.9\%$.
Both $K_\mathrm{eq}$ and $K_\tau$ provide good estimates of the
permeability for the lattice-like geometries, with MAPEs below $20\%$.
In contrast, $K_\mathrm{sb}$, based on the constant single-bar
diameter, is nearly independent of~$\alpha$ and yields a
MAPE of $80.2\%$, confirming that it cannot reproduce the
angle-dependent permeability.

For the channel-like geometries ($\alpha \leq \SI{10}{\degree}$), all three
models underestimate the permeability. These correlations are derived for
multiply connected pore structures and do not account for the aligned,
low-resistance void spaces present at small rotation angles.


The Ergun-based Forchheimer coefficients,
$C_\mathrm{F}^\mathrm{Erg}(d_\mathrm{sb})$ and
$C_\mathrm{F}^\mathrm{Erg}(d_\mathrm{eq})$, computed from
\cref{eq:CF-ergun} using~$K_\mathrm{sb}$ and~$K_\mathrm{eq}$
respectively, are compared with the DNS-fitted values in
\cref{fig:Cf_number}. Because \cref{eq:CF-ergun} relates~$C_\mathrm{F}$
directly to~$\sqrt{K}$, the quality of the permeability estimate
propagates into the predicted Forchheimer coefficient.

The single-bar estimate $C_\mathrm{F}^\mathrm{Erg}(d_\mathrm{sb})$ yields
a constant value of approximately~$0.7$, which underestimates the
DNS-fitted~$C_\mathrm{F}$ for the lattice-like geometries
($\alpha \geq \SI{15}{\degree}$). Since~$d_\mathrm{sb}$ does not vary with the
rotation angle, the resulting~$K_\mathrm{sb}$ cannot capture the
angle-dependent permeability, and consequently the predicted Forchheimer
coefficient inherits this limitation. The module-equivalent estimate
$C_\mathrm{F}^\mathrm{Erg}(d_\mathrm{eq})$ varies with~$\alpha$ through
the angle-dependent specific surface area and reaches values closer to the
DNS data at some intermediate rotation angles. However, it significantly
overestimates~$C_\mathrm{F}$ at small rotation angles
($\alpha \leq \SI{10}{\degree}$), where the channel-like flow structure departs
from the packed-bed assumptions underlying the Ergun correlation.
At $\alpha = \SI{90}{\degree}$, both Ergun estimates also
overestimate~$C_\mathrm{F}$, consistent with the low DNS-fitted value
discussed above, which is attributed to the delayed transition to the
inertial regime at this angle. Moreover, the DNS data are limited to $\mathrm{Re}_\mathrm{p} \leq 200$,
which may not be sufficient to fully establish the Forchheimer coefficient
for all geometries. The development of
geometry-aware models for the Forchheimer coefficient, supported by
simulations at higher Reynolds numbers, remains a subject
for future investigation.

\subsection{Transition to inertial regime}\label{sec:transition}

\begin{figure}[tb] \centering	\includegraphics[width=0.45\textwidth]{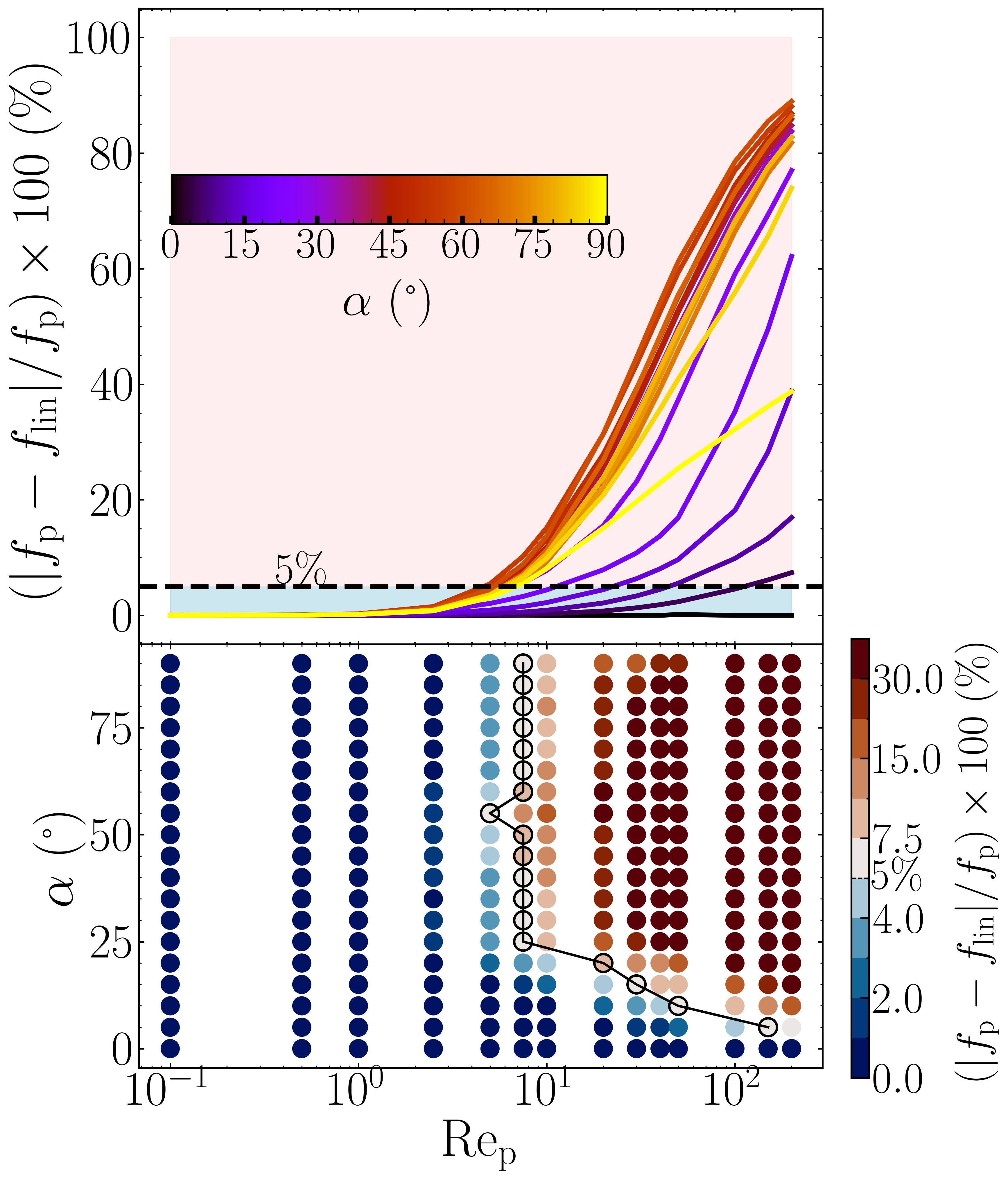} 
	\caption{Relative deviation of the friction factor from
	 the Darcy-regime value. The circular markers with black
	 outlines show the critical Reynolds number for each angle. } \label{fig:friction_factor_deviation}
\end{figure}

According to the classification proposed by \citet{Dybbs1984ANL},
this transition typically occurs near $\mathrm{Re}_\mathrm{p} \approx 1$
for spherical particles in packed beds. The present geometry
consists of square bars arranged at a constant rotation angle,
leading to void space connectivity characteristics that differ 
significantly from those of a sphere-packed medium. 

\citet{Hlushkou2006} suggested that if the inertial contribution exceeds $5\%$, the flow
should no longer be considered in the Darcy regime. The same criterion is adopted in the
present study. 
To identify the Reynolds number at which inertial effects become significant, the difference between 
the total friction factor and the linear friction factor 
($f_{\mathrm{lin}} = 1/(\phi\, \mathrm{Re}_{\mathrm{p}}\, \mathrm{Da})$) is computed 
and then divided by the total friction factor.
However, other criteria have 
also been reported in the literature, such as introduced by \citet{Zeng2006,bagci2014}. 
In future work, different transition criteria may be used to 
provide a more comprehensive assessment of the transition to inertia-dominated flow.

We illustrate the transition from viscosity to inertia-dominated regime
in \cref{fig:friction_factor_deviation}.
The upper figure shows 
the relative friction factor for each angle. As expected, the deviation remains 
zero for $\mathrm{Re}_\mathrm{p} \le 1$. With increasing $\mathrm{Re}_\mathrm{p}$,
the inertial contribution represented by the Forchheimer term grows with the Reynolds number. Consequently, 
the pressure drop no longer follows the linear Darcy regime, and the deviation from Darcy’s law increases rapidly.

At $\mathrm{Re}_\mathrm{p}=200$, the nonlinear component accounts for
80--90\% of the total friction factor for the fully lattice-like angles.
The channel-like configurations
($\alpha \leq \SI{10}{\degree}$) exhibit the lowest inertial contributions,
as the aligned flow passages suppress flow separation.
The angles $\alpha \in \{\SI{15}{\degree}\text{--}\SI{25}{\degree}\}$ and
$\{\SI{85}{\degree}\text{--}\SI{90}{\degree}\}$, although belonging to the
lattice-like group based on their bed-to-particle diameter ratio and permeability, also
show a reduced inertial contribution. At these angles, the void space connectivity promotes preferential flow
paths that give rise to localised jet-like structures, delaying the
transition to the inertial regime relative to the fully lattice-like
configurations.

The bottom figure displays the same dataset from a different
perspective. The blue regions represent viscous flow,
while the red areas indicate increasing influence of inertial 
effects. The transition to the inertial regime is not abrupt.
The figure shows a transitional region in which neither viscous
forces nor inertial forces fully dominate. Points at which the 
inertial contribution exceeds \SI{5}{\percent} are identified 
as the critical Reynolds number ($\mathrm{Re}_\mathrm{crit}$), 
following the criterion of \citet{Hlushkou2006}; these points
are marked with black circular symbols and connected for each angle.
For most geometries,
the critical Reynolds number is approximately 
$\mathrm{Re}_\mathrm{p} \approx 7.5$.  
The present study effectively captures the onset of the inertial
effects within the range $\mathrm{Re}_\mathrm{p} \le 200$. 
A detailed investigation of the fully developed inertial regime,
as well as finer simulations in the $\mathrm{Re}_\mathrm{p} = 50\text{--}100$
interval to refine the location of $\mathrm{Re}_\mathrm{crit}$, remain subjects for future work.

At $\alpha = \SI{55}{\degree}$, the transition to the inertial
regime occurs at approximately $\mathrm{Re}_\mathrm{p} \approx 5$,
which is lower than for the other geometries. 
Potential factors contributing to this behaviour include
the sensitivity of the Darcy number evaluation for this 
orientation and the unique connectivity of the void space, 
which may promote early localized recirculation.
A conclusive explanation would require a detailed flow field
analysis which currently exceeds the scope of this study.

\section{Conclusion}
\label{sec:Conclusion}

In this study, numerical simulations were performed using OpenFOAM-12
to investigate viscosity- and inertia-dominated flow through periodically
arranged square-bar structures over a range of $\mathrm{Re}_\mathrm{p}=0.1 \text{--} 200$
with different angles ($\SI{0}{\degree} \text{--} \SI{90}{\degree}$) at fixed porosity $\phi = 0.332$.
A total of 266 simulations were performed using a predominantly hexahedral mesh,
and selected cases were successfully validated against available PIV measurements obtained from 
the experimental configuration of \citet{christin2025}.
Some important conclusions can be drawn from the obtained results:

\begin{enumerate}[label=(\arabic*)]

  \item The rotation angle has a strong influence on the macroscopic flow
        characteristics.  For $\alpha \le \SI{10}{\degree}$, the structure forms
        \emph{channel-like} geometries, resulting in high permeability.  As the angle increases beyond $10^{\circ}$, the
        connectivity between successive layers becomes more complex, leading to the
        development of recirculation zones and increased frictional losses, resulting
        in heterogeneous velocity fields. We refer to these configurations
        ($\alpha \geq 15^{\circ}$, including $90^{\circ}$) as \emph{lattice-like}
        geometries. The $\alpha = \SI{90}{\degree}$ configuration is also
        classified as lattice-like, as confirmed by its Darcy number,
        bed-to-particle diameter ratio and tortuosity.

  \item  Using a 5\% inertial contribution threshold, the transition to the
        inertial regime occurs at $\mathrm{Re}_\mathrm{p} \approx 7.5$ for
        lattice-like geometries, with the earliest onset at
        $\alpha = \SI{55}{\degree}$.

  \item The drag characteristics are strongly impacted by the
        geometric configuration, and the angle yielding the highest
        friction factor shifts with the Reynolds number. In the
        viscous regime, the severe constriction of the flow path at
        $\alpha = \SI{25}{\degree}$ results in the highest friction
        factor. In the inertial regime, the contraction--expansion
        sequences at $\alpha = \SI{60}{\degree}$ lead to flow
        separation and recirculation, resulting in the highest
        friction factor.

 \item The classification into channel-like and lattice-like regimes can be
        supported quantitatively by the bed-to-particle diameter ratio and the hydraulic
        tortuosity. Geometries with $D/d_{\mathrm{eq}} < 3$ and $\tau < 1.1$
        behave as channel-like, whereas those with $D/d_{\mathrm{eq}} > 3$
        and $\tau > 1.1$ fall into the lattice-like group.

  \item The module-equivalent diameter $d_{\mathrm{eq}}$, which accounts for
        the angle-dependent wetted surface area, collapses the friction
        factor data onto the Ergun correlation more effectively than the
        constant single-bar diameter $d_{\mathrm{sb}}$, allowing to
        partially account for the geometry variation. Among the
        permeability models, the Blake--Kozeny estimate based
        on~$d_{\mathrm{eq}}$ provides the best agreement with the DNS data
        for the lattice-like geometries (MAPE~$= 12.3\%$). The
        tortuosity-based estimate $K_\tau$ (Kozeny constant $c_{\mathrm{KC}} = 3.1$)  reproduces the angle-dependent trend but with larger deviations
        (MAPE~$= 18.9\%$), while the single-bar estimate yields
        substantially larger errors (MAPE~$= 80.2\%$). All three models
        underestimate the permeability of the channel-like configurations
        ($\alpha \leq 10^{\circ}$), as they are formulated for multiply
        connected pore structures.

  \item The Forchheimer coefficient $C_\mathrm{F}$ varies strongly with
        rotation angle, peaking at $\alpha = 60^{\circ}$ and remaining low
        for the channel-like and $90^{\circ}$ configurations. At intermediate
        angles ($30^{\circ}$--$70^{\circ}$), $C_\mathrm{F}$ fluctuates most strongly, reflecting the sensitivity of inertial losses to
        the void-space topology in this range.

\end{enumerate}

The good agreement between simulations and
experiments confirms that the present approach provides a reliable
framework for analysing flow through such periodic structures.  Future
work should extend the analysis to higher Reynolds numbers to examine
the onset of unsteady and turbulent flow, and include additional
configurations to improve the fitting of geometry-aware models for
both permeability and the Forchheimer coefficient.

\section*{Acknowledgments}
The authors gratefully acknowledge the financial support of the Deutsche
Forschungsgemeinschaft (DFG) through No. SFB/ TRR287, Project No. 422037413.
The authors wish to thank Christin Velten, Kerstin Hülz, and Katharina Zähringer for sharing the dataset of the experimental measurements of flow in the packed bed.

\appendix
\setcounter{figure}{0}
\setcounter{table}{0}
\renewcommand{\thefigure}{A.\arabic{figure}}
\renewcommand{\thetable}{A.\arabic{table}}
\section{Effect of mesh resolution on the numerical results}
\label{app:VerificationMeshSensitivity}

\begin{figure}[tb] \centering
	\includegraphics[width=0.8\linewidth]{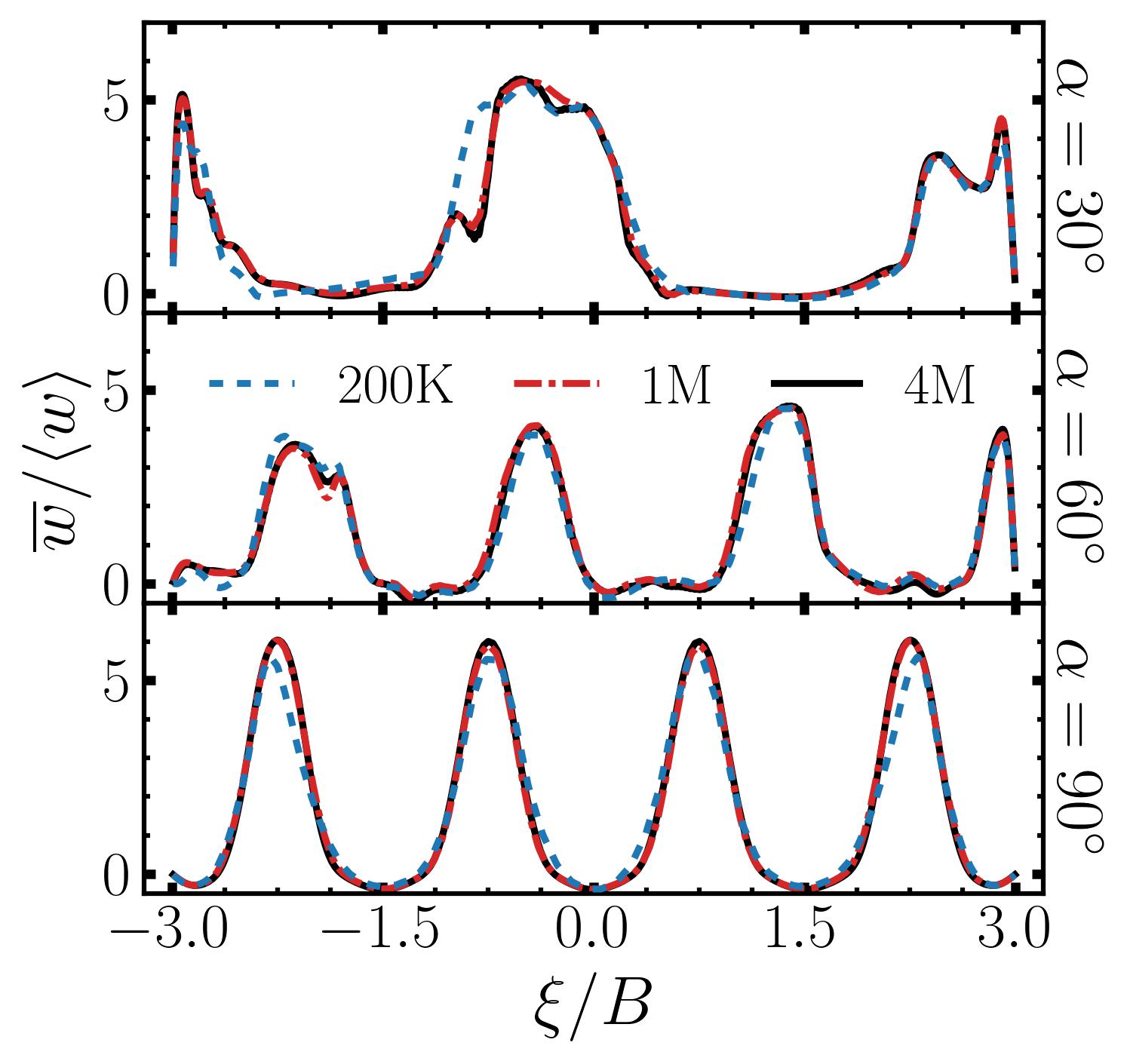} 
	\caption{Normalized spanwise-averaged velocity profiles at P3 (see \cref{fig:ModuleGeometry}) for angles $\SI{30}{\degree}$, $\SI{60}{\degree}$, and $\SI{90}{\degree}$. Results are shown for three mesh resolutions (200K, 1M, and 4M cells)}
	\label{fig:verification_POS3_Re200_}
\end{figure}

\begin{figure}[tb] \centering
	\includegraphics[width=0.8\linewidth]{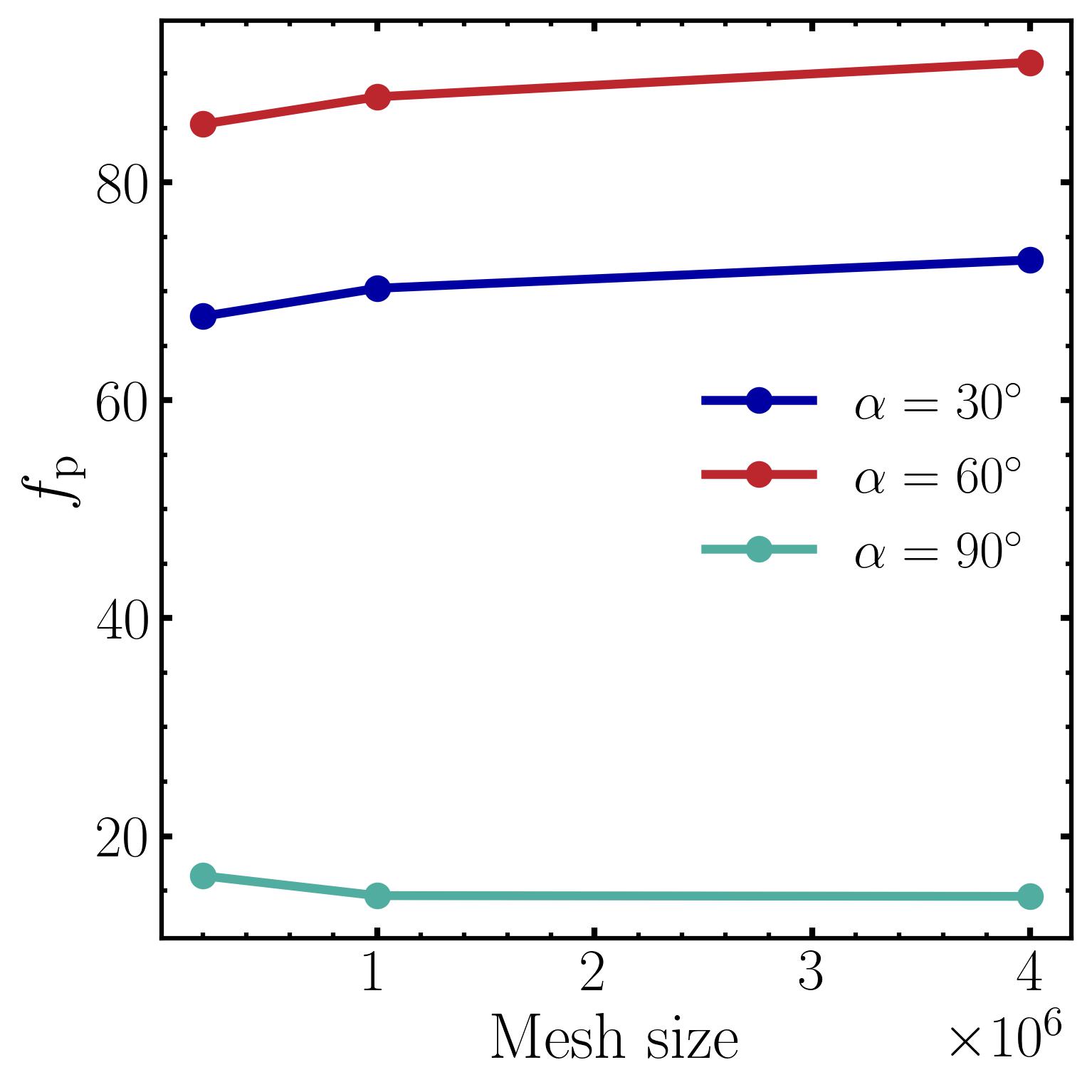} 
	\caption{Friction factor for three mesh sizes (200K, 1M, and 4M cells) at angles of $30^\circ$, $60^\circ$, and $90^\circ$. }
	\label{fig:Mesh_sensty_friction_factor}
\end{figure}

\begin{table}[tb]
	\centering
	\caption{Friction factors used in for each angle and mesh. The table contains the number of cells per layer
		($N_\mathrm{C}$)} 
	\label{tab:friction_factors}
	\begin{tabular}{c ccc}
		\toprule
		 \diagbox[width=1.5cm]{$\alpha$}{$N_\mathrm{C}$}  & \num{2e5} & \num{1e6} & \num{4e6} \\
		\midrule
		$\SI{30}{\degree}$ & 67.70 & 70.27 & 72.87 \\
		$\SI{60}{\degree}$ & 85.33 & 87.84 & 91.01 \\
		$\SI{90}{\degree}$ & 16.36 & 14.53 & 14.47 \\
		\bottomrule
	\end{tabular}
\end{table}

A mesh sensitivity analysis to the results was performed using three unrefined meshes 
containing approximately \num{2e5}, \num{1e6}, and \num{4e6} cells per module at 
$\mathrm{Re}_\mathrm{p}=200$. No local refinement was applied in the flow region.
 Three modules with different rotation angle, namely \(\alpha = \SI{30}{\degree}\),
\(\SI{60}{\degree}\), and \(\SI{90}{\degree}\), were considered. The objective of this analysis was to
assess the influence of the overall mesh resolution on the solution accuracy and to determine an
optimal mesh for the subsequent simulations.

The analysis was conducted for the case shown in \cref{fig:verification_POS3_Re200_},
focusing on the first module of the computational domain along the probe line located at
point P3 (see \cref{fig:ModuleGeometry}), positioned at the mid-line in the
\(z\)-direction. Velocity data were extracted from statistically steady-state solutions
without applying periodic averaging. \Cref{fig:verification_POS3_Re200_} presents the
normalized spanwise-averaged velocity profiles. The coarsest mesh (\num{2e5} cells)
shows a noticeable deviation from the finer ones, particularly in the central region of the
\(\SI{30}{\degree}\) case. For \(\SI{60}{\degree}\) and \(\SI{90}{\degree}\), the deviation
becomes evident near the peak locations of the velocity profiles, where the \num{2e5} cell mesh results in
less accurate results. As the mesh is refined, the profiles converge toward those of
the finest mesh, indicating improved resolution of the flow field.
For all investigated angles, the \num{1e6} and \num{4e6} meshes yield nearly
identical profiles, suggesting that the flow features are well-resolved from the \num{1e6} cell mesh.

Similarly, the friction factor results shown in \cref{fig:Mesh_sensty_friction_factor} confirm this
trend. \Cref{fig:Mesh_sensty_friction_factor} also presents the same angles, mesh resolutions and Reynolds number.
Increasing the cell number from \num{2e5} to \num{1e6} produces a noticeable change in $f_\mathrm{p}$,
while the difference between the \num{1e6} and \num{4e6} cases is minimal.
We can also see this in \cref{tab:friction_factors}, where the friction factor changes by
approximately \SI{3.7}{\percent}, \SI{3.6}{\percent}, and \SI{0.41}{\percent} for
\SI{30}{\degree}, \SI{60}{\degree}, and \SI{90}{\degree}, respectively, when increasing the
mesh size from \num{1e6} to \num{4e6}.
The mesh sensitivity analysis demonstrates that the
\num{1e6} cell mesh provides reliable results while maintaining a reasonable 
computational cost.

\clearpage
\bibliography{references}

\end{document}